\documentclass[aps,prb,superscriptaddress,onecolumn,preprintnumbers,amsmath,amssymb,11pt]{revtex4-1}
\usepackage{epsfig}
\usepackage{graphicx}
\usepackage{dcolumn}
\usepackage{subfigure}
\usepackage[bookmarks=false]{hyperref}
\hypersetup{colorlinks=true,
           citecolor=red,
            filecolor=blue,
           urlcolor=blue}
\usepackage{ulem}
\usepackage{bm}
\usepackage{color}
\usepackage{braket}
\usepackage{latexsym}
\usepackage{amsmath}
\usepackage{amssymb}
\usepackage{latexsym}

\def \beq{\begin{eqnarray}}
\def \eeq{\end{eqnarray}}
\def \r{{\mathbf{r}}}
\def \rp{{\mathbf{r}^{\prime}}}
\def \R{{\mathbf{R}}}
\def \Rp{{\mathbf{R}^{\prime}}}
\def \Q{{\mathbf{Q}}}
\def \S{{\mathbf{S}}}
\def \k{{\mathbf{k}}}
\def \K{{\mathbf{K}}}
\def \T{{\mathcal{T}}}

\def \xh{\hat{x}}
\def \yh{\hat{y}}

\begin{document}

\title{Superconductivity from a confinement transition out of a fractionalized Fermi liquid with $\mathbb{Z}_2$ topological and Ising-nematic orders}

\author{Shubhayu Chatterjee}
\affiliation{Department of Physics, Harvard University, Cambridge Massachusetts
02138, USA.}
\author{Yang Qi}
\affiliation{Institute for Advanced Study, Tsinghua University, Beijing 100084, China}
\affiliation{Perimeter Institute of Theoretical Physics, Waterloo Ontario-N2L 2Y5, Canada.}
\author{Subir Sachdev}
\affiliation{Department of Physics, Harvard University, Cambridge Massachusetts
02138, USA.}
\affiliation{Perimeter Institute of Theoretical Physics, Waterloo Ontario-N2L 2Y5, Canada.}
\author{Julia Steinberg}
\affiliation{Department of Physics, Harvard University, Cambridge Massachusetts
02138, USA.}

\date{\today \\
\vspace{1.6in}}
\begin{abstract}
The Schwinger-boson theory of the frustrated square lattice antiferromagnet yields a stable, gapped $\mathbb{Z}_2$ spin liquid ground 
state with time-reversal symmetry, incommensurate spin correlations and long-range Ising-nematic order. We obtain an equivalent description of this state using fermionic spinons (the fermionic spinons can be considered to be bound states of the bosonic spinons and the visons). Upon doping, the $\mathbb{Z}_2$ spin liquid can lead to a fractionalized Fermi liquid (FL*) with small Fermi pockets of electron-like quasiparticles, while preserving 
the $\mathbb{Z}_2$ topological and Ising-nematic orders. We describe a Higgs transition out of this deconfined metallic state into a confining superconducting state which is almost always of the Fulde-Ferrell-Larkin-Ovchinnikov type, with spatial modulation of the superconducting order.
\end{abstract}
\maketitle
\tableofcontents
\section{Introduction}

The $\mathbb{Z}_2$ spin liquid is the simplest gapped quantum state with time-reversal symmetry and bulk anyon
excitations \cite{ReadSachdev,RJSS91,WenSqLattice,MVSS99,TSMPAF00,RMSLS01,Kitaev03,Freedman04}. For application to the cuprate
superconductors, an attractive parent Mott insulating state is a $\mathbb{Z}_2$ spin liquid obtained in the
Schwinger boson mean field theory of the square lattice antiferromagnet with first, second, and third neighbor
exchange interactions \cite{ReadSachdev,ReadSachdev2,SubirQPT}. This is a fully gapped state with incommensurate
spin correlations, spinon excitations which carry spin $S=1/2$, vison excitations which carry $\mathbb{Z}_2$ magnetic
flux, and long-range Ising nematic order associated with a breaking of square lattice rotation symmetry. 
Upon doping away from such an insulator with a density of $p$ holes,
we can obtain a FL* metallic state which inherits the topological order of the $\mathbb{Z}_2$ spin liquid, and acquires a 
Fermi surface of electron-like quasiparticles enclosing a volume associated with a density of $p$ 
fermions \cite{SS94,XGWPAL96,FFL,XGW06,RK07,RK08,YQSS10,Mei12,MPSS12,DCSS14,Punk15,chowdhury2015enigma}.
It was also noted \cite{FFL} that a $\mathbb{Z}_2$-FL* metal can undergo a transition into a superconducting state which is concomitant
with confinement and the loss of $\mathbb{Z}_2$ topological order (while preserving the Ising-nematic order).
Given the recent experimental evidence for a Fermi-liquid-like metallic state in the underdoped cuprates with a density of $p$ positively 
charged carriers \cite{Marel13,MG14,LTCP15}, the present paper will investigate the structure of the confining superconducting
state which descends from the $\mathbb{Z}_2$-FL* state associated with Schwinger boson mean field theory of the 
square lattice \cite{ReadSachdev,ReadSachdev2,SubirQPT}.

For insulating $\mathbb{Z}_2$ spin liquids, the spectrum can be classified by four separate `topological' or 
`superselection' sectors, which are conventionally labeled $1$, $e$, $m$, and $\epsilon$ \cite{Kitaev03}. In
the Schwinger boson theory, the Schwinger boson itself becomes a bosonic, $S=1/2$ spinon excitation which we identify
as belonging to the $e$ sector. The vison, carrying $\mathbb{Z}_2$ magnetic flux, is spinless, and we label this as belonging
to the $m$ sector. A fusion of the bosonic spinon and a vison then leads to a fermionic spinon \cite{RC89}, which belongs
to the $\epsilon$ sector. We summarize these, and other, characteristics of insulating $\mathbb{Z}_2$ spin liquids
in Table~\ref{tab:z2}.
\begin{table}[h]
\begin{tabular}{| c || c | c | c | c || c | c | c | c ||}
\hline 
 & $1$ & $e$ & $m$ & $\epsilon$ & $1_c$ & $e_c$ & $m_c$ & $\epsilon_c$ \\
\hline \hline
$S$ & 0 & 1/2 & 0 & 1/2 & 1/2 & 0 & 1/2 & 0 \\
\hline
Statistics & boson & boson & boson & fermion & fermion & fermion & fermion & boson \\
\hline
Mutual semions & $-$ & $m$, $\epsilon$, $m_c$, $\epsilon_c$ & $e$, $\epsilon$, $e_c$, $\epsilon_c$ & $e$, $m$, $e_c$, $m_c$ &
 $-$ & $m$, $\epsilon$, $m_c$, $\epsilon_c$ & $e$, $\epsilon$, $e_c$, $\epsilon_c$ & $e$, $m$, $e_c$, $m_c$ \\
 \hline
 $Q$ & 0 & 0 & 0 & 0 & 1 & 1 & 1 & 1 \\
 \hline
 Field operator & $-$ & $b$ & $\phi$ & $f$ & $c$ & $-$ & $-$ & $B$ \\
\hline
\end{tabular}
\caption{Table of characteristics of sectors of the spectrum of the $\mathbb{Z}_2$-FL* state.
The first four columns are the familiar sectors of an insulating spin liquid. The value of $S$ indicates integer or half-integer representations
of the SU(2) spin-rotation symmetry. The ``mutual semion'' row lists the particles which have mutual seminionic statistics with the particle
labelling the column. The electromagnetic charge is $Q$. The last four columns 
represent $Q=1$ sectors present in $\mathbb{Z}_2$-FL*, and these are obtained by adding an electron-like quasiparticle, $1_c$, 
to the first four sectors. The bottom row denotes the fields operators used in the present paper to annihilate/create particles in the sectors.}
\label{tab:z2}
\end{table}

For a metallic $\mathbb{Z}_2$-FL* state, it is convenient to augment the insulating classification by counting the charge, $Q$, of
fermionic electron-like quasiparticles: we simply add a spectator electron, $c$, to each insulator sector, and label the resulting states
as $1_c$, $e_c$, $m_c$, and $\epsilon_c$, as shown in Table~\ref{tab:z2}. It is a dynamical question of whether the $c$ particle
will actually form a bound state with the $e$, $m$, or $\epsilon$ particle, and this needs to be addressed specifically for each
Hamiltonian of interest.

Now let us consider a confining phase transition in which the $\mathbb{Z}_2$ topological order is destroyed.
This can happen by the condensation of one of the non-trivial bosonic particles of the $\mathbb{Z}_2$-FL* state.
From Table~\ref{tab:z2}, we observe that there are three distinct possibilities:
\begin{enumerate}
\item Condensation of $m$: this was initially discussed in Refs.~\onlinecite{RJSS91,MVSS99}. For the case of insulating
antiferromagnets with an odd number of $S=1/2$ spins per unit cell, the non-trivial space group transformations of the $m$
particle lead to bond density wave order in the confining phase. The generalization to the metallic $\mathbb{Z}_2$-FL* state
was presented recently in Ref.~\onlinecite{PCAS16}.  
\item Condensation of $e$: now we are condensing a boson with $S=1/2$, and this leads to long-range antiferromagnetic order \cite{CSS94,TGTS09,Kaul08,SWHSL16,WS16}.
\item Condensation of $\epsilon_c$: this is a boson which carries electromagnetic charge, and so the confining state is a superconductor \cite{FFL}.
\end{enumerate}

This paper will focus on the third possibility listed above: condensation of $\epsilon_c$, the bosonic ``chargon''. Our specific interest is in the 
Schwinger boson $\mathbb{Z}_2$ spin liquid of Refs.~\onlinecite{ReadSachdev,ReadSachdev2,SubirQPT}. 
To study the $\epsilon_c$ states in this model, we need to consider the fusion of the $\epsilon$ quasiparticle
and the electron (which is in the $1_c$ sector). Thus a key ingredient needed for our analysis will be the projective transformations
of the $\epsilon$ particle under the symmetry group of the underlying square lattice antiferromagnet. 
These transformations are not directly available from the Schwinger boson mean-field theory, which is expressed in terms
of the $e$ boson. However, remarkable recent advances \cite{EH13,Unification_Luetal,QiFu,ZLV15,Zheng_Tri,Lu_Triangular,WangSqLattice} 
have shown how the projective symmetry group (PSG) of the $\epsilon$ particle can be computed from a knowledge of the PSG of the $e$ and $m$ particles. 

Section~\ref{sec:mapping} describes in detail our computation of the PSG of the $\epsilon$ excitations of the square
lattice Schwinger boson $\mathbb{Z}_2$ spin liquid state. These results are then applied in Section~\ref{sec:sc}
to deduce the structure of the superconductor obtained by condensing $\epsilon_c$.

\section{Mapping between bosonic and fermionic spin liquids on the rectangular lattice via symmetry fractionalization}
\label{sec:mapping}

The Schwinger boson mean-field $\mathbb{Z}_2$ spin-liquid described in  Refs.~\onlinecite{ReadSachdev,ReadSachdev2,SubirQPT} spontaneously breaks the $C_4$ rotation symmetry of the square lattice, and this nematic order persists in the $\mathbb{Z}_2$-FL*. Therefore, we identify the space-group symmetries of the rectangular lattice along with time reversal $\T$ as the symmetries that act projectively on the $e$ and $m$ particles (bosonic spinons and visons respectively) in the above ansatz in the Schwinger boson representation (bSR). Below, we briefly describe the idea of symmetry fractionalization \cite{EH13,Unification_Luetal,QiFu,ZLV15,Zheng_Tri,Lu_Triangular,WangSqLattice}, which enables us to find the projective actions of the same symmetries on the $\epsilon$ particles, or equivalently the spinons in the Abrikosov fermion representation (fSR). We only provide a quick summary, and refer the reader to the references above for detailed discussions.  

The key idea behind symmetry fractionalization is that the action of any symmetry on a physical state (which must necessarily contain an even number of any anyon in a $\mathbb{Z}_2$ spin liquid) can be factorized into local symmetry operations on each of these anyons. For concreteness, consider the translation operator $T_x$ ($T_y$), which translates the wave-function by one unit in the $\xh$ ($\yh$) direction, and a physical state $\ket{\psi}$ that contains two $e$ particles at $\r$ and $\rp$. We assume that this operation can be factorized as:
\beq
T_x \ket{\psi} = T^e_x(\r) T^e_x(\rp) \ket{\psi}
\eeq
Since the $e$ particle is coupled to emergent gauge fields, $T^{e}_x(\r)$ is not invariant under gauge transformations. But if we consider a set of operations that combine to the identity, $T^e_x T^e_y (T^e_x)^{-1} (T^e_y)^{-1}$ for example, then the combined phase that the $e$ particle picks up is gauge-invariant.  In a gapped $\mathbb{Z}_2$ spin liquid, this phase must be $\pm1$. This can be seen by fusing two $e$ particles, which is a local excitation and therefore can only pick up a trivial phase $+1$. This also implies that this phase is independent of location of the $e$ particle as long as translation symmetry is preserved by the spin liquid. Although we chose the $e$ particle for illustration, an analogous picture holds for $m$ and $\epsilon$ particles as well. 

Generalizing this to other symmetries including internal ones like time-reversal $\T$, we can find a quantized gauge invariant phase of $\pm 1$ for each series of symmetry operations that combine to identity on the physical wave-function. This phase is fixed for a given anyon in a particular spin liquid, and is also referred to as the symmetry fractionalization quantum number. These quantum numbers are universal features of $\mathbb{Z}_2$ spin liquids, and provide a way to characterize topological order without parton constructions. However, given a particular parton construction (either in terms of bosons or fermions), we can determine these quantum numbers --- we shall illustrate how to so for the particular bosonic $\mathbb{Z}_2$ spin liquid we are interested in. Also, given a set of quantum numbers we can attempt to find a corresponding spin liquid ansatz --- we again explicitly describe this later when we find a fermionic mean-field ansatz. But first, we outline how we find these quantum numbers for the fermions from those of the bosons and the visons. 

In a $\mathbb{Z}_2$ spin liquid, the $e$ and $m$ particle satisfy the following fusion rule \cite{Kitaev03}:
\beq
e \times m = \epsilon
\eeq
In other words, we can think of the fermionic spinon ($\epsilon$) as a bound state of the bosonic spinon ($e$) and the vison ($m$). Therefore, in most cases, for a set of symmetry operations $O$ combining to identity, the phase factor picked up by the fermionic spinon $\sigma^{\epsilon}_{O}$ is just the product of the phase $\sigma^{e}_{O}$ picked up by the bosonic spinon and the phase $\sigma^{m}_{O}$ picked up by the vison. These have been referred to as the trivial fusion rules in Ref.~\onlinecite{Unification_Luetal}. In certain cases, there is an additional factor of $-1$ coming from the non-trivial mutual statistics between the spinon and the vison, and these fusion rules are called non-trivial. Once these fusion rules are known, the symmetry fractionalization quantum numbers for the $\epsilon$ can be calculated from those of $e$ and $m$. 

With this preamble, we now outline the procedure to derive the fermionic spin liquid ansatz corresponding to the bosonic $\mathbb{Z}_2$ spin liquid obtained from the $J_1$-$J_2$-$J_3$ antiferromagnetic Hamiltonian on the square lattice \cite{ReadSachdev,ReadSachdev2}. We first describe the symmetries of the spin liquid, and list the elementary combinations for which we need to calculate the symmetry fractionalization quantum numbers. Then we discuss the idea of PSG for the Schwinger boson spin liquids in general  \cite{WangVishwanath}, and use it to calculate the afore-mentioned quantum numbers for our bosonic ansatz. We proceed with analogous derivations of the quantum numbers for the visons \cite{YHMPSS11,KSYHYB15} and fermions \cite{WenSqLattice,LuRanLee_PRB2013,LeeNagaosaWen,BieriPRB16} using PSG techniques. We then derive the non-trivial fusion rules, and use these to relate the bosonic and fermionic symmetry quantum numbers of time-reversal preserving mean-field spin liquids on the rectangular lattice. Finally, we find the specific set of quantum numbers for the fermionic spin liquid of our interest, and find an ansatz consistent with this particular pattern of symmetry fractionalization. 

\subsection{Symmetries of the spin liquid}
Consider a mean-field Hamiltonian with the following symmetries: global spin-rotations, action of the rectangular lattice space group and time-reveral $\T$. Since a mean-field spin liquid ansatz is explicitly invariant under global SU(2) spin-rotations, we only need to consider the projective actions of the other symmetries. Let us define the lattice points $\r$ = $x \, \xh + y \, \yh = (x,y)$ in a rectangular coordinate system with unit vectors $\xh$ and $\yh$. The space group of the rectangular lattice is then generated by the translations and reflections $\in$ $\lbrace T_{x}, T_{y}, P_{x}, P_{y}  \rbrace$, defined as follows:
\begin{subequations}
\begin{eqnarray}
T_{x}: & (x,y) \rightarrow (x+1, y) \\
T_{y}: & (x,y) \rightarrow (x, y+1) \\
P_{x}: & (x,y) \rightarrow (-x, y) \\
P_{y}: & (x,y) \rightarrow (x, -y) 
\end{eqnarray}
\end{subequations}
There are algebraic constraints which relate these generators. Below, we present the finite set of elementary combinations of these generators that are equivalent to the identity operator on any physical wave-function.
\begin{subequations}
\label{crysSymbegin}
\beq
T_x^{-1} T_y^{-1} T_x T_y, ~
P_x^2, ~
P_y^2,  ~
P_x^{-1} T_x P_x T_x, ~
P_x^{-1} T_y^{-1} P_x T_y, ~
P_y^{-1} T_x^{-1}P_y T_x, ~
P_y^{-1} T_y P_y T_y \text{ and }
P_x^{-1}P_y^{-1}P_x P_y  \nonumber \\
\label{crysSymend}
\eeq
When we include time-reversal $\T$, we also have to consider the following additional operators:
\beq
\mathcal{T}^2,
T_x^{-1} \mathcal{T}^{-1} T_x \T, ~
T_y^{-1} \mathcal{T}^{-1} T_y \T, ~
P_x^{-1} \mathcal{T}^{-1} P_x \T \text{ and } 
P_y^{-1} \mathcal{T}^{-1} P_y \T
\eeq
\end{subequations}
These are the combinations for which we need to calculate the symmetry fractionalization quantum numbers, and all other combinations that lead to identities can be expressed as products of these elementary combinations.  
 
\subsection{PSG for bSR}
\subsubsection{Schwinger boson ansatz}
The spin operator can be represented in terms of Schwinger bosons operators $b_{\r \alpha}$  as
\begin{equation}
\vec{S}_{\r}= \frac{1}{2} b^{\dagger}_{\r \alpha} \vec{\sigma}_{\alpha \beta} b_{\r \beta}
\end{equation}
where $\alpha= \uparrow, \downarrow$.
The mean field Hamiltonian is
\begin{equation}
\label{eq:bsrmfh}
H^{b}_{MF}=-\sum_{\r \rp}(Q_{\r \rp}\epsilon_{\alpha \beta}b^{\dagger}_{\r \alpha}b^{\dagger}_{\rp \alpha}+ H.c.)+\sum_{\r}\lambda_{\r} ( b^{\dagger}_{\r \alpha}b_{\r \alpha} - 1)
\end{equation}
where $\lambda_{\r}$ is a Lagrange multiplier that enforces the single occupancy constraint $\sum_{\alpha}b^{\dagger}_{\r \alpha}b_{\r \alpha}=1$ on an average and the  $Q_{\r \rp} = \langle \epsilon_{\alpha \beta} b_{\r \alpha}b_{\rp \beta} \rangle $ are mean-field pairing link variables that satisfy $Q_{\r \rp}= -Q_{\rp \r}$.  
The Schwinger boson SL wavefunction is
\begin{equation}
\label{eq:bsrwf}
\vert \Psi^{b} \rangle=P_{G}\enspace \text{exp}\left[\sum_{\r \rp} \xi_{\r \rp}\epsilon_{\alpha \beta} b^{\dagger}_{\r \alpha}b^{\dagger}_{\rp \beta}\right]\vert 0 \rangle
\end{equation}
where $P_G$ projects onto states with a single spin, and $\xi_{\r \rp} = - \xi_{\rp \r}$ is obtained by diagonalizing $H^{b}_{MF}$ via a Bogoliubov transformation. 

\subsubsection{Gauge freedom, PSG and algebraic constraints}
Here, we formally introduce the PSG in the context of the Schwinger bosons, and describe its relation to the symmetry fractionalization quantum numbers. This discussion closely follows Ref.~\onlinecite{WangVishwanath}. In the bSR, consider the following local $U(1)$ transformation of the bosons:
\begin{equation}
\label{eq:1}
b_{\textbf{r} \alpha} \rightarrow e^{i\phi(\textbf{r})}b_{\textbf{r}\alpha}
\end{equation}
This leaves all the physical observables unchanged, but the mean field ansatz undergoes the following transformation to leave the Hamiltonian invariant:
\begin{equation}
Q_{\r \rp} \rightarrow e^{i\phi(\r)+i\phi(\rp)} Q_{\r \rp}
\end{equation}
Any two mean field ansatz that are related by a local U(1) transformation as described above correspond to the same physical wave function after projection to single spin-occupancy per site. Therefore, a spin liquid state has a particular symmetry $X$ if the corresponding mean field ansatz is invariant under the symmetry action of $X$ followed by an additional local gauge transformation $G_X$. 
\beq
\label{eq:2}
G_{X}: b_{\textbf{r} \alpha} &\rightarrow& e^{i\phi_{X}(\textbf{r})}b_{\textbf{r}\alpha} \nonumber \\
G_X X: Q_{\r \rp} &\rightarrow& \text{exp}\left[ i (\phi_X[X(\r)] +\phi_X[X(\rp)]) \right] Q_{X(\r)X(\rp)}
\eeq
The set of all such transformations $\{G_X X\}$ that leave the ansatz invariant form the PSG. Ideally, each PSG element should reflect a physical symmetry of the ansatz. But it turns out that there are certain transformations in the PSG that are not associated with any physical symmetry, but still leave the ansatz invariant. In other words, these are purely local transformations, and correspond to the identity operation $X=\mathbb{I}$. They form a subgroup of the PSG, called the invariant gauge group (IGG) \cite{WenSqLattice}. It is natural to associate these members of the PSG with the emergent gauge field in the spin liquid. For $\mathbb{Z}_2$ spin liquids, the IGG is therefore $\mathbb{Z}_2$, generated by $-1$.
 
One can now ask: how is the IGG related to the $\mathbb{Z}_2$ symmetry fractionalization quantum numbers? To answer this question, note that elements of the IGG correspond to identity transformations on the ansatz, and therefore on the physical wave-function as well (assuming that the mean-field state survives projection). Therefore, for any series of operations that combine to the identity, the corresponding projective operation should be an element of the IGG [for example, for $T_x^{-1} T_y T_x T_y^{-1} = \mathbb{I}$, we have $ (G_{T_x}T_{x} )^{-1} (G_{T_y}T_{y})(G_{T_x}T_{x})(G_{T_Y}T_{y})^{-1} = \pm 1$]. At the same time, note from Eqs.~(\ref{eq:1}) and 
(\ref{eq:2}) that this projective operation describes the gauge-invariant phase that a single $e$ particle picks up under this set of transformations. Therefore, the element of the IGG which we choose for a spin liquid ansatz is precisely the symmetry fractionalization $\mathbb{Z}_2$ quantum number for this set of operations. In other words, the symmetry fractionalization quantum numbers determine the particular extension of the physical symmetry group by the IGG that is realized by a given spin liquid.

The algebraic relations between the spatial symmetry operations in a group strongly constrain the possible choices of gauge transformations $G_X$ associated with symmetry operations $X$. Without referring to a particular ansatz, we can use these relations to find all possible PSGs for a set of symmetries. Below, we find the most general phases $\phi_X$ consistent with the algebraic constraints on a rectangular lattice with time-reversal symmetry. 

\subsubsection{Solutions to the algebraic PSG}
We just state the solutions here, and present the derivation in Appendix \ref{bosPSG}. The solutions for the phases $\phi_X$ (modulo 2$\pi$), as defined in Eq.~(\ref{eq:2}) can be written down in terms of integers $\{ p_i \}$ defined modulo 2, which are precisely the symmetry fractionalization quantum numbers for the $e$ particles in the spin liquid. 
\begin{subequations}
\label{eq:bpsg}
\begin{eqnarray}
\phi_{T_{x}}(x,y) &=& 0 \\
\phi_{T_y}(x,y) &=& p_1 \pi x \\
\phi_{P_x}(x,y) &=& p_2 \pi x + p_4 \pi y + \frac{p_6}{2} \pi \\
\phi_{P_y}(x,y) &=& p_3 \pi x + p_5 \pi y + \frac{p_7}{2} \pi \\
\phi_{\mathcal{T}}(x,y) & = & p_8 \pi x + p_9 \pi y 
\end{eqnarray}
\end{subequations}

\subsubsection{PSG solutions for the nematic spin liquid ansatz for the $J_1$-$J_2$-$J_3$ model on the square lattice}
We need to find the quantum numbers for the Schwinger boson mean-field ansatz of our interest, which is given by  \cite{ReadSachdev2,SubirQPT}:
\begin{equation}
Q_{i,i+ \hat{x}} \neq Q_{i,i+ \hat{y}} \neq 0, Q_{i,i + \hat{x} + \hat{y}} = Q_{i,i-\hat{x} + \hat{y}} \neq 0, Q_{i,i+2\hat{x}} \neq 0, Q_{i,i + 2 \hat{y}} = 0 
\end{equation}
All the mean-field variables are real in a particular gauge choice, so time-reversal symmetry is preserved. 
This state has nematic order, as the following gauge-invariant observable $\mathcal{I} = |Q_{i,i+ \hat{x}}|^2 - |Q_{i,i+ \hat{y}}|^2 \neq 0$.
This state has the following solution for $\{p_i\}$, which we can derive (as shown in Appendix \ref{BosAnsPSG}) by using the transformation of the ansatz under the symmetry operation $X$ to fix the phases $\phi_X$ (or correspondingly, the integers $p_i$):
\begin{equation}
p_1 = 0, p_2 = 0, p_3 = 0, p_4 = 1, p_5 = 1, p_6 = 1, p_7 = 0, p_8 = 0, p_9 = 0
\end{equation}

\subsection{Vison PSG}

\begin{figure}[!h]
\begin{center}
\includegraphics[scale = 0.6]{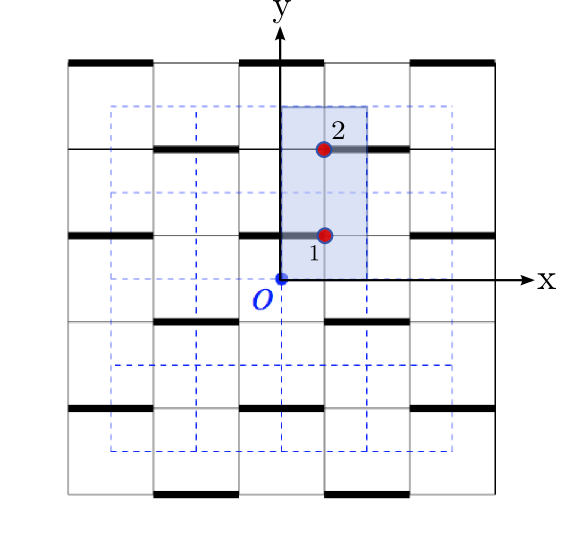}
\caption{(Color online) The gauge choice for $J_{\R \Rp}$ on the rectangular lattice. The dark and light bonds respectively represent links with $J_{\R \Rp} = -1$ and $J_{\R \Rp} = 1$. The unit cell is denoted by the blue box, and the sub lattice indices by 1 and 2. Dotted blue lines form the original lattice.}
\label{FFIMgauge}
\end{center}
\end{figure}

In this section, we shall derive the vison PSG for the rectangular lattice. To do so, we shall resort to a description of the visons by the fully frustrated transverse field Ising model on the dual lattice \cite{BalentsCenke_VBS}. Denoting the points on the dual lattice by $\R$, the vison Hamiltonian is given by 
\begin{equation}
H = \sum_{ \R \Rp} J_{\R \Rp} \, \tau^{z}_{\R} \tau^{z}_{\Rp} - \sum_{\R} h_{\R} \, \tau^{x}_{\R}
\end{equation}
where the product of bonds around each elementary plaquette ($\Box$) is negative, given by
\begin{equation}
\prod_{\Box} \text{sgn}(J_{\R \Rp}) = -1
\end{equation}
Note that this Hamiltonian is invariant under the gauge transformation 
\begin{equation}
\tau^{z}_{\R} \rightarrow \eta_{\R} \, \tau^{z}_{\r}, \; \; J_{\R \Rp} \rightarrow \eta_{\R} \, \eta_{\Rp} \, J_{\R \Rp}, \; \; \eta_{\R} \in \{ \pm1\}  = \mathbb{Z}_2
\end{equation}
For calculating the vison PSG, we make the following gauge choice (depicted in Fig. \ref{FFIMgauge}):
\beq
J_{\R, \R + \xh} = (-1)^{x+y} =  J_{\R + \xh, \R} \; \text{ and } J_{\R, \R + \yh} = 1 = J_{\R + \yh, \R} 
\eeq

Let us consider the spatial symmetry generators first. Since the Hamiltonian is invariant under symmetry transformations only upto a gauge transformation, we identify, for each symmetry generator $X$ in the space group of the rectangular lattice, an element $G_X \in \mathbb{Z}_2$ such that 
\beq
G_X  X[ J_{\R \Rp}] = J_{X[\R]X[\Rp]} G_X[X(\R)]G_X[X(\Rp)] =  J_{\R \Rp}
\eeq
Note that all operations are defined with respect to the original lattice. From Fig. \ref{FFIMgauge}, we can immediately see what the required gauge transformations are. Since the $x$ bonds change sign under $T_x, T_y$ and $P_y$, whereas the $y$ bonds are invariant, we must have $G_{T_x} = G_{T_y} = G_{P_y} = (-1)^{X}$. Further, $P_{x}$ acts trivially on both the $x$ and $y$ bonds, so $G_{P_x} =1$. 
Now, consider time-reversal $\T$. Since the Ising couplings $J_{\R \Rp} = \pm 1$ are real, these are invariant under $\T$, so $G_{\T} = 1$ as well. With this knowledge of additional phases under lattice transformations, we can calculate the symmetry fractionalization quantum numbers of the visons in a manner analogous to the bosons --- we list these in Table \ref{RelationTable} under the column $\sigma^{m}_{O}$. 

We comment that these are exactly the quantum numbers one would obtain by thinking of the vison acquiring an extra phase of $-1$ when it is transported adiabatically with $\pi$-flux per unit cell, corresponding to an odd number of spinons. The results are also consistent with another calculation from a soft-spin formulation of the visons, which we present in Appendix 
\ref{visPSG}.

\subsection{PSG for fSR}
\subsubsection{Schwinger fermion ansatz}
In terms of fermion operators, the spin operator $\S_{\r}$ can be written as
\begin{equation}
\S_{\r}=\frac{1}{2}f^{\dagger}_{\r \alpha} \vec{\sigma}_{\alpha \beta} f_{\r \beta}
\end{equation} 
We write down the Hamiltonian in terms of two different mean fields as follows \cite{LeeNagaosaWen}:
\begin{eqnarray}
H^{f}_{MF} &=& \sum_{\r \rp}\frac{3}{8}J_{\r \rp} \left[ \chi_{\r \rp} f^{\dagger}_{\r,\alpha}f_{\rp,\alpha} + \Delta^f_{\r \rp}\epsilon_{\alpha \beta} f^{\dagger}_{\r,\alpha} f^{\dagger}_{\rp,\beta} + H.c - |\chi_{\r \rp}|^2 - |\Delta^f_{\r \rp}|^2  \right] 
\nonumber \\
&& + \sum_{\r} a_0^{3}(f^{\dagger}_{\r \alpha}f_{\r \alpha} -1) + [(a_0^{1} + i a_0^{2})\epsilon_{\alpha \beta} f^{\dagger}_{\r \alpha} f^{\dagger}_{\r \beta} + H.c ]
\end{eqnarray}
where we have defined the spinon hopping amplitude $\chi_{\r \rp} \delta_{\alpha \beta}$ and the spinon-pairing amplitude $\Delta^f_{\r \rp}\epsilon_{\alpha \beta}$, both spin-rotation invariant (and non-zero in general), as follows:
\begin{eqnarray}
\Delta^f_{\r \rp}\epsilon_{\alpha \beta} &=& -2\langle f_{\r \alpha}f_{\rp \beta}\rangle, \enspace \Delta^f_{\r \rp} =\Delta^f_{\rp \r},
\\
\chi_{\r \rp}\delta_{\alpha \beta} &=& 2\langle f^{\dagger}_{\r \alpha}f_{\rp \beta}\rangle, \enspace \chi_{\r \rp} = \chi^{*}_{\rp \r},
\end{eqnarray}
and we have also introduced Lagrange multipliers $a_0^{i}$ to enforce single occupancy per site on average. 

\subsubsection{Gauge freedom, PSG and algebraic constraints}
In order to see the local SU(2) symmetry of the Hamiltonian, let us introduce 
\begin{equation}
\psi_{\r}=\begin{pmatrix}
\psi_{1 \r}
\\
\psi_{2 \r}
\end{pmatrix}=\begin{pmatrix}
f_{\r \uparrow}
\\
f^{\dagger}_{\r \downarrow}
\end{pmatrix}
\end{equation}
We also define a mean-field matrix $U_{\r \rp}$ as follows:
\begin{equation}
\label{Umatrix}
U_{\r \rp}=\begin{pmatrix}
\chi^{*}_{\r \rp} & \Delta_{\r \rp}
\\
\Delta^{*}_{\r \rp} & -\chi_{\r \rp}
\end{pmatrix}=U^{\dagger}_{\rp \r}
\end{equation}
In terms of the $\psi$ fermions, the single occupancy constraints reduce to $
\langle \psi^{\dagger}_{\r}\tau^{l} \psi_{\r} \rangle =0$, so the mean field Hamiltonian can now be written as
\begin{equation}
H^{f}_{MF}=\sum_{\r \rp}\frac{3}{8}J_{\r \rp}\left[\frac{1}{2}\text{Tr}(U^{\dagger}_{\r \rp}U_{\r \rp})-\psi^{\dagger}_{\r}U_{\r \rp}\psi_{\r}+h.c)\right]+\sum_{\r} a^{l}_{0}(\r) \psi^{\dagger}_{\r}\tau^{l}\psi_{\r}
\end{equation}
Note that $U_{\r \rp}$ is not a member of SU(2) as det$(U)<0$, but $i U_{\r \rp} \in$ SU(2) up to a normalization constant. $H^{f}_{MF}$ is explicitly invariant under a local SU(2) gauge transformation $W(\r)$:
\begin{subequations}
\begin{align}
\psi_{\r} \rightarrow W(\r)\psi
\\
U_{\r \rp} \rightarrow W(\r)U_{\r \rp}W^{\dagger}(\rp)
\end{align}
\end{subequations} 
In general, dynamical SU(2) gauge fluctuations can reduce the gauge group. In particular, in presence of non-collinear SU(2) flux, the SU(2) gauge bosons become massive and the only the $\mathbb{Z}_2$ gauge structure is unbroken at low energies \cite{WenSqLattice,LeeNagaosaWen}. In the following sections, we shall only consider $\mathbb{Z}_2$ as the IGG, generated by $-\tau^{0}$. 

Analogous to the bosonic case, we define the PSG as the set of all transformations (symmetry transformations followed by gauge transformations) that leave the ansatz $U_{\r \rp}$ invariant (this will also leave the $a_0^l$s invariant as these are self-consistently determined by the $U_{\r \rp}$s). Pure gauge fluctuations, corresponding to the identity element in the physical symmetry group, make up the IGG. Hence operators in the symmetry group that combine to the identity in the physical group, can only be $\pm \tau^{0} \in$ IGG in the projective representation. Similar to the bosonic case, this element $\eta = \pm \mathbb{I}$ will determine the symmetry fractionalization quantum number for the corresponding series of operations. 

\subsubsection{Solutions to the algebraic PSG}
Algebraic relations between the symmetry group [\ref{crysSymbegin}] elements will lead to a series of conditions for the gauge transformations $G_X[\r]$, which are now SU(2) valued. The general solutions (without referring to any ansatz) are given below in terms of $\mathbb{Z}_2$ valued variables $\{\eta \}$, and derived in Appendix \ref{fermPSG}.  
\begin{subequations}
\begin{align}
G_{T_{x}}(x,y) &= \tau^{0}
\\
G_{T_{y}}(x,y) &= (\eta_{T_xT_y})^{x} \tau^{0}
\\
G_{P_x}(x,y) &= ( \eta_{P_x T_x})^{x} (\eta_{P_x T_y})^{y} g_{P_x}, \; g_{P_x} \in SU(2), \; g_{P_x}^{2} = \eta_{P_x} \tau^{0} 
\\
G_{P_y}(x,y) &= ( \eta_{P_y T_x})^{x} (\eta_{P_y T_y})^{y} g_{P_y}, \; g_{P_y} \in SU(2), \; g_{P_y}^{2} = \eta_{P_y} \tau^{0} 
\\
G_{\T}(x,y) &= (\eta_{\T T_x})^x (\eta_{\T T_y})^y g_{\T}, \; g_{\T} \in SU(2), \; g_{\T}^{2} = \eta_{\T} \tau^{0} 
\end{align}
\end{subequations}
where the SU(2) matrices are bound by the following constraints:
\beq
g_{P_x} g_{\T} g_{P_x}^{-1} g_{\T}^{-1} = \eta_{\T P_x} \tau^{0}, \; 
g_{P_y} g_{\T} g_{P_y}^{-1} g_{\T}^{-1} = \eta_{\T P_y} \tau^{0}, \;
g_{P_x} g_{P_y} g_{P_x}^{-1} g_{P_y}^{-1} = \eta_{P_x P_y} \tau^{0}  
\eeq

\subsection{Fusion rules}

We provide a table for trivial and non-trivial fusion rules for $\mathbb{Z}_2$ spin liquids on the rectangular lattice with time reversal symmetry $\T$, and provide proofs/arguments in the Appendix \ref{FusRules}.

\begin{equation}
\begin{tabular}{| l | r |}
\hline 
\text{Commutation relation} & \text{Fusion rule} \\
\hline \hline
$T_x^{-1} T_y^{-1} T_x T_y$ & Trivial \\
$ P_x^2 $ & Non-trivial\\
$P_y^2 $ & Non-trivial \\ 
$P_x^{-1} T_x P_x T_x$ & Trivial \\
$P_x^{-1} T_y^{-1} P_x T_y $ & Trivial \\
$P_y^{-1} T_x^{-1}P_y T_x$ & Trivial \\
$P_y^{-1} T_y P_y T_y$ & Trivial \\
$P_x^{-1}P_y^{-1}P_x P_y$ & Non-trivial \\
$\mathcal{T}^2$ & Trivial \\
$T_x^{-1} \mathcal{T}^{-1} T_x \mathcal{T}$ & Trivial \\
$T_y^{-1} \mathcal{T}^{-1} T_y \mathcal{T}$ &  Trivial \\
$P_x^{-1} \mathcal{T}^{-1} P_x \mathcal{T}$ & Non-trivial \\
$P_y^{-1} \mathcal{T}^{-1} P_y \mathcal{T}$ & Non-trivial \\
\hline
\end{tabular}
\end{equation}

\subsection{Fermionic ansatz}
\subsubsection{General relation between bosonic and fermionic PSGs for rectangular lattice}
In Table \ref{RelationTable}, we use the anyon fusion rules to relate bosonic symmetry fractionalization quantum number $\sigma^{e}_{O}$ with the fermionic one $\sigma^{\epsilon}_{O}$ for $\mathbb{Z}_2$ spin liquids. These are related as follows:
\beq
\sigma^{\epsilon}_{O} =  \sigma^{t}_{O} \sigma^{e}_{O} \sigma^{m}_{O}
\eeq
where we have used the knowledge of the vison quantum number $\sigma^{m}_{O}$, and the twist factor $ \sigma^{t}_{O}$ which is $-1$ for non-trivial fusion rules and $+1$ otherwise.

\begingroup
\renewcommand*{\arraystretch}{1.5}
\begin{table}[h]
\begin{tabular}{| l | c | c | c | c | r |}
\hline 
\text{Commutation relation} & $\sigma^{e}_{O}$   & $\sigma^{\epsilon}_{O}$ &  $\sigma^{m}_{O}$  & $\sigma^{t}_{O}$ & \text{Relation} \\
\hline \hline
$T_x^{-1} T_y^{-1} T_x T_y$ & $ (-1)^{p_1} $ & $ \eta_{T_x T_y} $ & -1 & 1 & $ (-1)^{p_1 + 1} = \eta_{T_x T_y} $\\
$P_x^{-1} T_x P_x T_x$ & $ (-1)^{p_2} $ & $ \eta_{P_x T_y} $ & 1 & 1 & $ (-1)^{p_2} =  \eta_{P_x T_x}$ \\
$P_y^{-1} T_x^{-1}P_y T_x$ & $ (-1)^{p_3} $ & $ \eta_{P_y T_x} $ & -1 & 1 & $ (-1)^{p_3 + 1} =  \eta_{P_y T_x}$  \\
$P_x^{-1} T_y^{-1} P_x T_y $ & $ (-1)^{p_4} $ & $ \eta_{P_x T_y} $ & -1 & 1 & $ (-1)^{p_4 + 1} =  \eta_{P_x T_y}$  \\
$P_y^{-1} T_y P_y T_y$  & $ (-1)^{p_5} $ & $ \eta_{P_y T_y} $  & 1 & 1 & $ (-1)^{p_5} =  \eta_{P_y T_y}$ \\
$ P_x^2 $ & $ (-1)^{p_6} $ & $ \eta_{P_x} $  & 1 & -1 & $ (-1)^{p_6 + 1} =  \eta_{P_x} $ \\
$P_y^2 $ & $ (-1)^{p_7} $ & $ \eta_{P_y} $ & 1 & -1 & $ (-1)^{p_7 + 1} =  \eta_{P_y} $  \\ 
$P_x^{-1}P_y^{-1}P_x P_y$ & 1 & $ \eta_{P_xP_y} $  & -1 & -1 & $1 = \eta_{P_xP_y} $ \\
$\mathcal{T}^2$ & -1 & -1 & 1 & 1 & 1 = 1\\
$T_x^{-1} \mathcal{T}^{-1} T_x \mathcal{T}$ & $(-1)^{p_8}$ & $\eta_{\mathcal{T}T_x }$ & 1 & 1 & $(-1)^{p_8} = \eta_{\mathcal{T}T_x }$ \\
$T_y^{-1} \mathcal{T}^{-1} T_y \mathcal{T}$ & $(-1)^{p_9}$ & $\eta_{\mathcal{T}T_y }$ & 1 & 1 & $(-1)^{p_9} = \eta_{\mathcal{T}T_y}$ \\
$P_x^{-1} \mathcal{T}^{-1} P_x \mathcal{T}$ & $(-1)^{p_6}$ & $\eta_{\mathcal{T}P_x }$ & 1 & -1 & $(-1)^{p_6+ 1} = \eta_{\mathcal{T}P_x }$ \\
$P_y^{-1} \mathcal{T}^{-1} P_y \mathcal{T}$ & $(-1)^{p_7}$ & $\eta_{\mathcal{T}P_y }$ & 1 & -1 & $(-1)^{p_7+1} = \eta_{\mathcal{T}P_y }$ \\
\hline
\end{tabular}
\caption{Correspondence between bosonic and fermionic $\mathbb{Z}_2$ spin liquids on a rectangular lattice with time-reversal symmetry $\T$}
\label{RelationTable}
\end{table}
\endgroup

\subsubsection{Specific fermionic ansatz}
Plugging in the values of $\{p_i\}$ for the bosonic ansatz in Table \ref{RelationTable}, we can find the desired values of $\eta_{XY}$s for the fermionic ansatz. Doing so and solving the matrix equations (details in Appendix \ref{Ferm1}), we find the following solutions for the $G_X$s:
\begin{subequations}
\label{FermGX}
\begin{eqnarray}
G_{T_x}(x,y) &=& \tau^{0}, \\
G_{T_y}(x,y) &=& (-1)^{x} \tau^{0}, \\
G_{P_x}(x,y) &=&  \tau^{0}, \\
G_{P_y}(x,y) &=& (-1)^{x + y} i \tau^{3}, \\
G_{\T}(x,y) &=& i \tau^{2} .
\end{eqnarray}
\end{subequations}
Now we solve for the allowed nearest-neighbor (NN), next-NN (NNN), and NNNN bonds demanding $G_X X(U_{\r \rp}) = U_{\r \rp}$ for each bond. The solution is an ansatz with $\pi$-flux through elementary plaquettes, with real pairing on the NN and NNN bonds, and real hopping on the NNNN bonds:
\begin{subequations}
\label{Fansatz1}
\begin{eqnarray}
U_{\r,\r+\hat{x}} &=& (-1)^{y} \Delta_{1x} \, \tau^{1}, \\
U_{\r,\r + \hat{y}} &=& \Delta_{1y} \, \tau^{1} , \\
U_{\r,\r + \hat{x} + \hat{y}} = U_{\r,\r - \hat{x} + \hat{y}} &=& (-1)^{y} \Delta_{2} \, \tau^{1} ,  \\
U_{\r, \r + 2\hat{x}} &=& - t_{2x} \tau^{3}, \\
U_{\r, \r + 2\hat{y}} &=& - t_{2y} \tau^{3} .
\end{eqnarray}
\end{subequations}
We note that this PSG also allows for an on-site chemical potential of the form $a^{3}_{0} \tau^{3}$, so that the density of fermions can be adjusted. An alternate derivation of the PSG of this fermionic ansatz, based on mapping of projected mean-field wave-functions, is presented in Appendix \ref{alt-fPSG} and serves as a consistency check for our results.

We can diagonalize the mean-field Hamiltonian corresponding to this using a two-site unit cell in the $y$-direction. Let $A$ and $B$ be the sublattice indices for $y$ even and odd respectively, and the reduced Brillouin zone (BZ) be given by $-\pi < k_x \leq \pi, -\pi/2 < k_y \leq \pi/2$. Since the up-spin and down-spin sectors decouple, we get a pair of degenerate bands. The Hamiltonian can be written in terms of a four-component Nambu-spinor $\Psi_{\k}$ as $H = \sum_{\k \in BZ} \; \Psi^{\dagger}_{\k} \; h(\k) \; \Psi_{\k}$, where 
\beq
\Psi_{\k} = \begin{pmatrix}
f_{\k A \uparrow} \\
f_{\k B \uparrow} \\
f^{\dagger}_{-\k A \downarrow} \\
f^{\dagger}_{-\k B \downarrow} 
\end{pmatrix}, \nonumber 
 \eeq
and $h(\k)$  is the 4 $\times$ 4 matrix given below in terms of  $ \varepsilon_{2\k} = - 2t_{2x} \text{cos}(2k_x) - 2t_{2y} \text{cos}(2k_y)$,
 \beq
  \begin{pmatrix}  \varepsilon_{2\k}   & 0 & 2 \Delta_{1x} \text{cos}(k_x) & 2\Delta_{1y}\text{cos}(k_y) \\
 &  & & + 4 i \Delta_2 \, \text{cos}(k_x) \text{sin}(k_y) \\
0 & \varepsilon_{2\k} & 2\Delta_{1y}\text{cos}(k_y) & -2 \Delta_{1x} \text{cos}(k_x) \\
& & - 4 i \Delta_2 \, \text{cos}(k_x)  \text{sin}(k_y) & \\
2 \Delta_{1x} \text{cos}(k_x) &  2\Delta_{1y}\text{cos}(k_y)  & -  \varepsilon_{2\k} & 0 \\ 
& +  4 i \Delta_2 \, \text{cos}(k_x)  \text{sin}(k_y) &  & \\
 2\Delta_{1y}\text{cos}(k_y) &- 2 \Delta_{1x} \text{cos}(k_x) & 0 & -  \varepsilon_{2\k} \\
 - 4 i \Delta_2 \, \text{cos}(k_x)  \text{sin}(k_y)  &  & & \\
 \end{pmatrix} \nonumber \\
 \eeq
 Diagonalizing this matrix gives us the spinon dispersion, with two doubly degenerate bands, \begin{equation}
E^{\pm}_{\k} = \pm \sqrt{ \left( 2t_{2x} \text{cos}(2k_x) + 2t_{2y} \text{cos}(2k_y) \right)^2 + 4 \left( \Delta_{1x}^2 \text{cos}^2(k_x) + \Delta_{1y}^2 \text{cos}^2(k_y)\right) + 16 \Delta_2^2 \text{cos}^2(k_x) \text{sin}^2(k_y)  }
\end{equation}

\begin{figure}[!htbp]
\begin{center}
\includegraphics[scale=0.47]{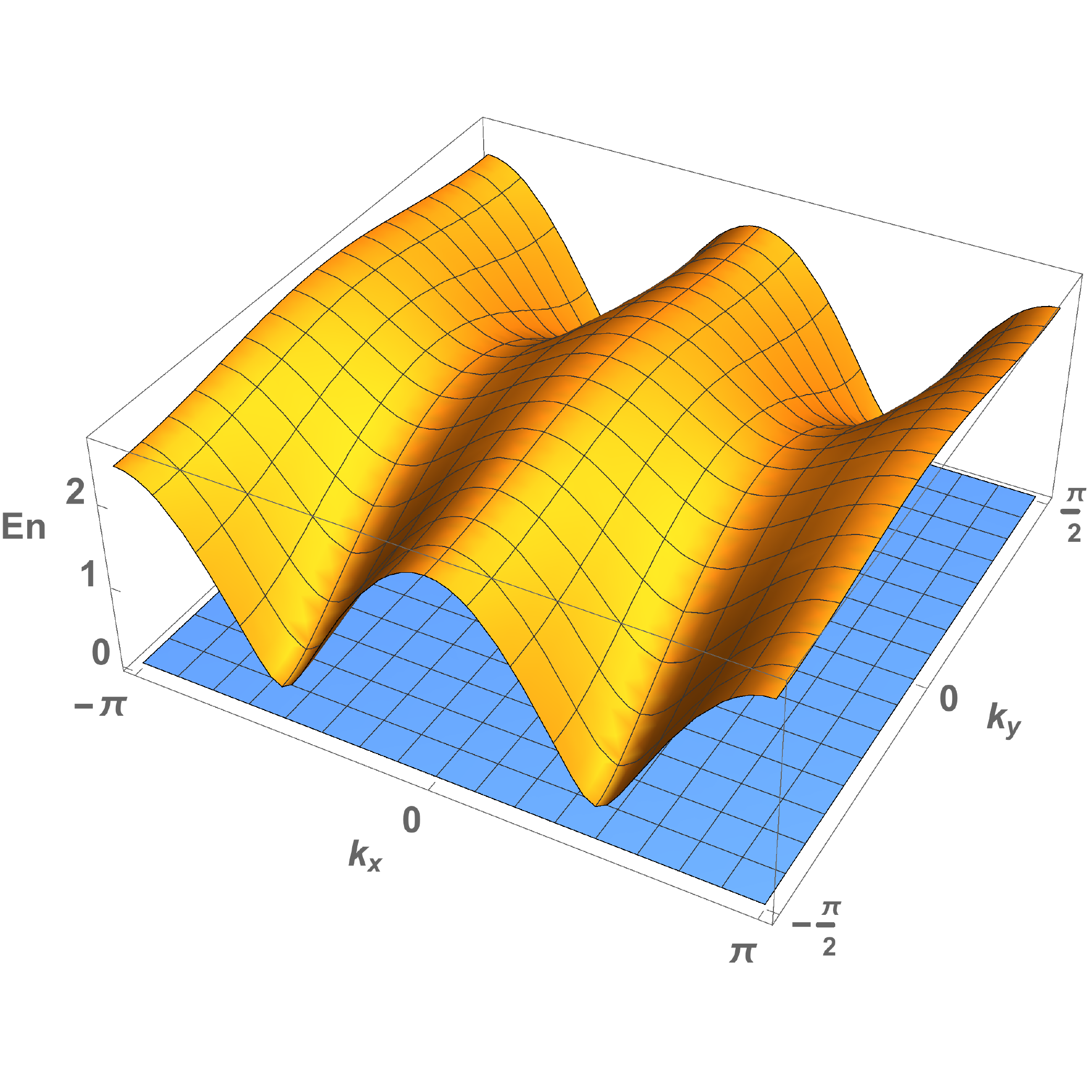}
\caption{(Color online) Mean field dispersion $E^{+}(\k)$ of the fermionic spinons for the parameters $(\Delta_{1x}, \Delta_{1y}, \Delta_2, t_{2x},t_{2y}) = (0.9,1,0.4,0.2,0.2)$. The other band is not shown for clarity.}
\label{dispers}
\end{center}
\end{figure}

Both these bands are fully gapped, with the mininum gap occurring at $(k_x,k_y) = (\pm \pi/2, \pm \pi/2)$ for $\Delta_{1x}, \Delta_{1y} \gg \Delta_2 \gg t_{2x}, t_{2y} $. $E^{+}_{\k}$ for typical parameter values is plotted in Fig. \ref{dispers}. 

Previous PSG studies have investigated fermionic spin liquids with space group symmetries of the square \cite{WenSqLattice,LeeNagaosaWen}, triangular and kagome \cite{LuRanLee_PRB2013,BieriPRB16} lattices, whereas we focus on the rectangular lattice. Reference~\onlinecite{FanHongVBS2012} discusses projected mean-field wave-functions of nematic spin liquids on the square lattice and their corresponding fermionic versions, but our initial bosonic state does not correspond to any of these states (as one can check by calculating fluxes through triangular plaquettes). We discuss the connection of their results with our work in greater detail in Appendix \ref{alt-fPSG}.

\section{Superconducting transition of the FL*}
\label{sec:sc}

So far, we have described the fermionic spinon excitations of the $\mathbb{Z}_2$ spin liquid.
These correspond to states in the $\epsilon$ sector of Table~\ref{tab:z2}. The $\mathbb{Z}_2$ FL* state has in addition
fermionic electron-like gauge-neutral excitations which belong the $1_c$ sector of Table~\ref{tab:z2}. These can be described
by some convenient dispersion for electron-like operators $c_{\k \sigma}$. In the recent analysis of Ref.~\onlinecite{Punk15}, 
the $c_{\k \sigma}$ states
were built out of electron orbitals which were centered on the bonds of the square lattice; on the other hand in Ref.~\onlinecite{YQSS10},
the $c_{\k \sigma}$ were obtained from electron-like states on the sites of the square lattice. The details of the dispersion and
Fermi surface structure of the $c_{\k \sigma}$ quasiparticles of the $\mathbb{Z}_2$-FL* will not be important here, and so we
simply assume they are characterized by some generic dispersion $\xi_{\k}$, and can be Fourier-transformed to operators $c_{\r \sigma}$
on the sites of the square lattice. Furthermore, the $c_{\r \sigma}$, being gauge-neutral, must have a trivial PSG.

Now we are interested in undergoing a confinement transition in which a boson, $B$, from the $\epsilon_c$ sector of 
Table~\ref{tab:z2} condenses. Such a boson is obtained by the fusion of the $\epsilon$ and $1_c$ states of Table~\ref{tab:z2}.
So we introduce two Bose operators on the sites of the square lattice transforming as
\beq
B_{1\r} \sim  c^{\dagger}_{\r \sigma} f_{r \sigma} \quad , \quad B_{2\r} \sim  \epsilon_{\sigma \sigma^{\prime}} c_{\r \sigma} f_{\r \sigma^{\prime}}.
\eeq
 Each of these bosonic operators carry a $\mathbb{Z}_2$ gauge charge of the $f$ fermions, and a U(1) charge corresponding to the $c$ fermions. We can then write down an effective Hamiltonian for the interplay between the $\epsilon$, $1_c$, and $\epsilon_c$ sectors
of Table~\ref{tab:z2}:
\beq
H  =  H_c + H^{MF}_f - \frac{J_K}{4} \sum_{\r, \rp} B^{\dagger}_{1\r}  c^{\dagger}_{\r \sigma} f_{r \sigma} + B^{\dagger}_{2\r}  \epsilon_{\sigma \sigma^{\prime}} c_{\r \sigma} f_{\r \sigma^{\prime}} + \mbox{H.c.}, \text{ where } \nonumber \\
H_c = \sum_{\k,\sigma} \xi_{\k} c^{\dagger}_{\k \sigma}  c_{\k \sigma} \text{ , and }  H^{MF}_f = \sum_{\r \rp , \sigma} \chi_{\r \rp} f^{\dagger}_{\r \sigma}  f_{\rp \sigma} + \sum_{\r \rp , \alpha \beta} \Delta^f_{\r \rp} \epsilon_{\alpha \beta}  f^{\dagger}_{\r \alpha}  f^{\dagger}_{\rp \beta} + \mbox{H.c.},
\eeq
where $J_K$ is the allowed `Kondo' coupling linking the sectors of $\mathbb{Z}_2$ FL* together.
A large $N$ approach, based on generalization of SU(2) to SU(N) yields only the term involving $B_{1\r}$ \cite{CS_PRB69,ReadNewns_JPC83}, but we consider a more simplistic mean-field approach where both bosons are present. At the transition, both these bosons condense together \cite{FFL}, and this leads to confinement. In the mean-field approximation, we replace $B_{i \r} = \langle B_{i \r} \rangle$ which is non-zero in the confined phase. The confinement transition out of this FL* state leads to a superconducting state \cite{FFL}, because a pairing between the spinons $f$ induces a pairing between the physical $c$ fermions when $\langle B_{i \r} \rangle \neq 0$. Further suppression of this superconductivity (by doping/magnetic field) will lead to a normal Fermi liquid state. Since the spin liquid ansatz breaks lattice symmetries, the confined states can also exhibit a density wave order. In the following subsection, we first detail the possible superconducting phases and describe how we obtain them from an effective bosonic Hamiltonian. 

\subsection{Possible confined phases}
On transition out of the FL*, we typically find that the superconducting phase is of the Fulde-Ferrell-Larkin-Ovchinnikov (FFLO) type \cite{FF_PR1964,LO_JETP1965}. This is a superconductor with fermion pairing only at finite momentum $\Q$, i.e., with spatial modulation of the order parameter $\Delta^{c}(\r) \sim e^{i \Q \cdot \r}$. It has also been referred to in the literature as a pair-density wave (PDW) state \cite{EBEFSKJT_NJP2009,BFK10,PAL_PRX2014,WAC14,WAC15,EFSKJT_RMP2015}. A PDW is distinct from a state with co-existing superconductivity and charge density wave (CDW) order. In particular, the superconducting order parameter has no uniform component, i.e, $\Delta_{\Q=0} = 0$; the Cooper pairs always carry a net momentum $\Q$.

In principle we can also have translation symmetry breaking in the particle-hole channel, leading to a generalized charge density wave order, often leading to oscillations of charge density on the bonds (a bond density wave). Following Ref.~\onlinecite{SachdevLaPlaca}, let us define a generalized density wave order parameter $P_{\Q_l}(\k)$ as
\beq
\langle c_{\r \sigma}^{\dagger} c_{\rp \sigma} \rangle = \sum_{\Q_l} \left( \int \frac{d^2k}{4\pi^2} P_{\Q_l}(\k) e^{i \k\cdot(\r - \rp)} \right) e^{i \Q_l\cdot(\r + \rp)/2}
\eeq
When $P_{\Q_l}(\k)$ is independent of $\k$, then the order parameter refers to on-site charge density oscillations at momentum $\Q_l$. When $P_{\Q_l}(\k)$ depends on $\k$, then it denotes charge density oscillations on the bonds, which is also often called a bond density wave \cite{SachdevLaPlaca}.

Note that a PDW at momentum $\Q$ typically leads to a CDW at momentum $\K = 2\Q$ \cite{EBEFSKJT_NJP2009}. This can be seen from a Landau-Ginzburg effective Hamiltonian, where a linear term in the CDW order parameter $P_{2\Q}$, of the form of $ \gamma_{\Delta}(\Delta_{\Q}^{*} \Delta_{-\Q} P_{2\Q} + \mbox{c.c})$ is allowed by symmetry. Therefore, in the phase where $\Delta_{\Q}$ is condensed, the system can always lower its energy by choosing a non-zero value of $P_{2\Q}$. Explicit computations later will show that boson condensation at finite momenta can lead to density wave states which have momenta different from $2 \Q_{PDW}$. These are therefore states where a PDW co-exists along with additional density wave order(s). 

To figure out the details of this transition at the level of mean-field theory, we first write down an effective Hamiltonian for the bosons $H_B$. This is determined by the PSG of the $f$ fermions, as described in Eq.~(\ref{FermGX}). Once we write down the effective Hamiltonian based on the PSG, we can find the minima of the boson dispersion at a set of momenta $\{ \Q_{i} \}$, at which the boson will condense on tuning to the phase transition. Across the transition, we can replace $B_{i \r}$ by the value of the condensate. The spinon-pairing $\Delta^f_{\r \rp}$ induces a pairing $ \Delta^c_{\r \rp}$ between the $c$ fermions, which is given in terms of the boson condensate by (perturbatively, to lowest non-zero order in $B_{i\r}$):
\beq
 \epsilon_{\alpha \beta} \Delta^c_{\r \rp}  = \langle \epsilon_{\alpha \beta}  c_{\r \alpha} c_{\rp \beta} \rangle \sim ( B_{1\r} B_{1\rp} + B_{2\r} B_{2\rp} ) \langle  \epsilon_{\alpha \beta}  f_{\r \alpha} f_{\rp \beta} \rangle = ( B_{1\r} B_{1\rp} + B_{2\r} B_{2\rp} )  \epsilon_{\alpha \beta} \Delta^f_{\r \rp}
\eeq

We also want to study if there is some density wave order, present on top of superconductivity or a PDW state. Therefore, in the confined phase we evaluate the order parameter $P_{\Q}(\k)$ by noting that 
\beq
\langle c_{\r \sigma}^{\dagger} c_{\rp \sigma} \rangle  \sim (B^{*}_{1\r} B_{1\rp} +  B^{*}_{2\r} B_{2\rp})\langle f_{\r \sigma}^{\dagger} f_{\rp \sigma} \rangle 
\eeq

Since each boson is a spin-singlet bound state of the $c$ and $f$ spinon, it has the same spatial symmetry fractionalization quantum numbers as the $f$ fermions. Time-reversal $\mathcal{T}$ interchanges $B_{1\r}$ and $B_{2\r}$ because of extra gauge transformation $G_{\tau}$ associated with the $f$ spinon. To deal with both bosons in a compact way, let us define a two-component spinor as follows:
\beq
B_{\r} = \begin{pmatrix}
B_{1\r} \\ B_{2\r}
\end{pmatrix}
\eeq
The action of the symmetry operations on $B_{\r}$ is derived in Appendix \ref{siteBosPSG}, here we just state the main results. Under any spatial symmetry operation $X_s$, this column vector just picks up an overall U(1) phase, because the gauge transformations $G_{X_s}$ for the $f$ fermions are all diagonal. 
\beq
\label{BosonPhases}
G_{X_s} X_s\left[ B_{\r} \right] = e^{i \phi_{X_s}[X_s(\r)]} B_{X_s[\r]}, \text{ with } \phi_{T_x} = 0, \, \phi_{T_y} = \pi x, \,  \phi_{P_x} = 0, \, \phi_{P_y} = \pi\left(x + y + \frac{1}{2} \right)
\eeq
However, time-reversal $\mathcal{T}$ mixes the up and down spinon operators, and imposes extra constraints. We demand $G_{X} X (H_{B}) = H_{B}$ for all symmetry operations $X$. Based on this, we can write down an effective Hamiltonian for the bosons as follows consistent with the PSG. For simplicity, we include only a $2\times 2$ hopping matrix $T_{\r \rp}$ upto next next nearest neighbors (we neglect pairing of bosons). We find that
\beq
\label{Hb}
H_b &=& \sum_{\r \rp} B^{\dagger}_{\r} \, T_{\r \rp} B_{\rp} + \mbox{H.c.} , \text{ where } T_{\r \rp} = T^d_{\r \rp} \tau^{0} + T^{od}_{\r \rp} \tau^{1}
\eeq
where $T^d$ and $T^{od}$ are the diagonal ($B_{1} \rightarrow B_{1}$ or $B_{2} \rightarrow B_{2}$) and off-diagonal ($B_1 \leftrightarrow B_2$) hopping elements, as described in Appendix \ref{siteBosPSG}. The diagonal hopping amplitudes are given by 
\beq
T^{d}_{\r, \r + \xh} = 0 , T^{d}_{\r, \r + \yh} = i T^{d}_{y}, T^{d}_{\r, \r + \xh + \yh} =  T^{d}_{\r, \r - \xh + \yh} = i T^{d}_{x+y}(-1)^y, T^{d}_{\r, \r + 2\xh} = T^{d}_{2x}, T^{d}_{\r, \r + 2\yh} = T^{d}_{2y}
\eeq
where all the $T^{d}_{\alpha}$ are real. The off-diagonal hopping is also exactly analogous, as the projective U(1) phases for both the $B_1$ and $B_2$ bosons are identical. However the overall coefficients $T^{od}_{\alpha}$ are not fixed by the PSG and generically different from $T^d_{\alpha}$. 

\begin{figure}[!htbp]
\includegraphics[scale=0.7]{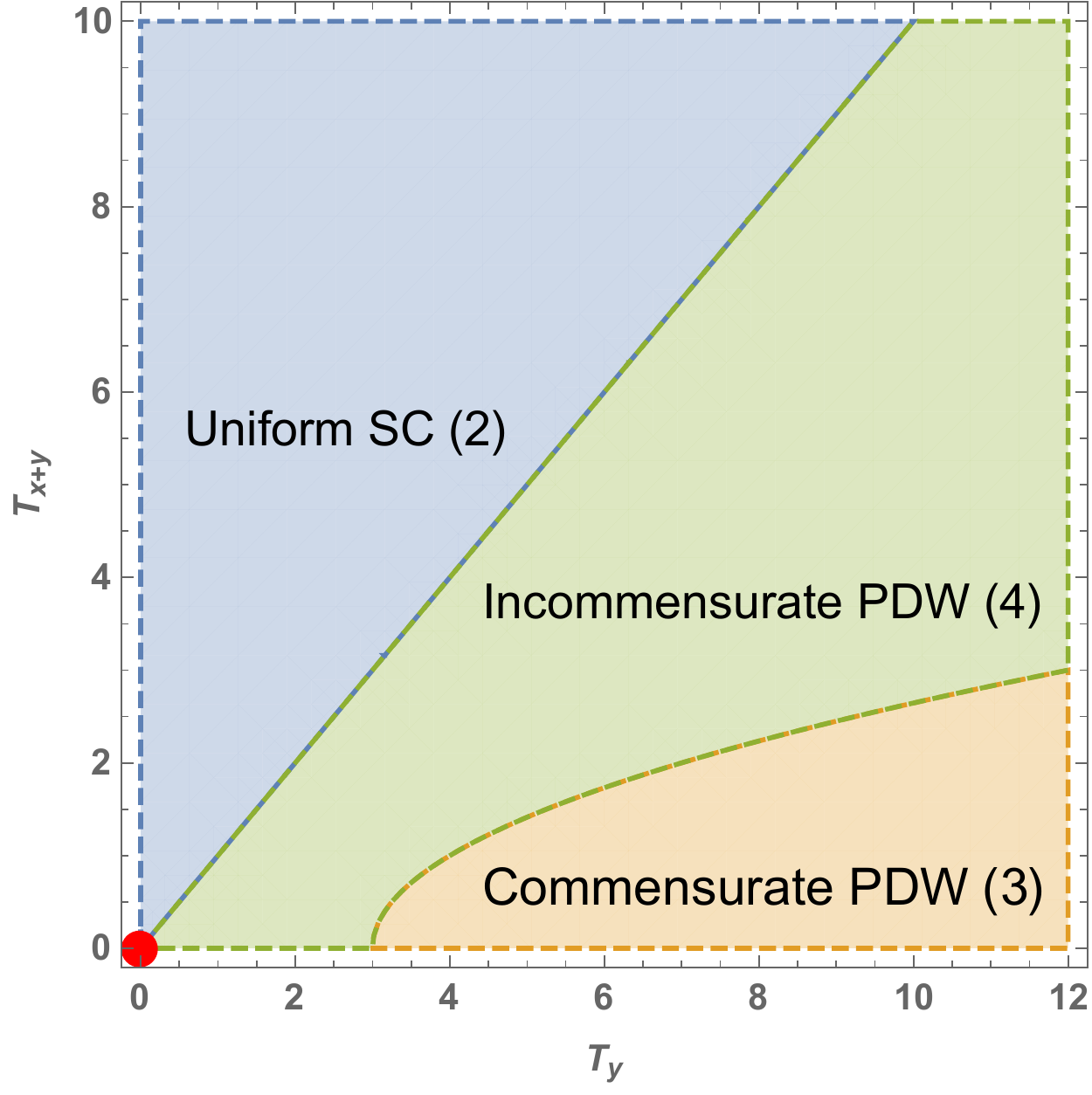}
\caption{(Color online) Phases of the superconductor; phase boundaries are approximate. $T_{2x}, T_{2y}$ are assumed small but non-zero. 
The number in brackets denotes the subsection in which the phase is discussed. 
The red dot denotes phase (1), a PDW state with unbroken $\T$. The phases are described in detail in the main text.}
\label{phaseDia}
\end{figure}

For simplicity, we first set the off-diagonal components $T^{od}_{\alpha}$ to zero by hand, which implies that we need to study only one boson --- let us call that $\mathcal{B}_{\r}$. We shall later argue that the resulting superconducting phases are essentially unchanged when one includes the off-diagonal components as well. Translational symmetry breaking in this gauge choice leads to an enlarged two-site unit cell in the $\hat{y}$ direction. Letting $A,B$ be the sublattice indices (for even/odd $y$), we define the Fourier transformed operators as 
\beq
\mathcal{B}_{\r \alpha} = \frac{1}{\sqrt{N_c}} \sum_{\k} e^{i \k\cdot\r_{\alpha}} \mathcal{B}_{\k\alpha}, ~~~ \alpha = A,  B
\eeq
where $N_c$ is the number of unit cells, and $- \pi < k_x \leq \pi, -\pi/2 < k_y \leq \pi/2$ defines the reduced BZ. Let us define $\Psi^{\dagger}_{\k} = (\mathcal{B}^{\dagger}_{\k A}, \mathcal{B}^{\dagger}_{\k B})$, then we can write $H_{\mathcal{B}} =  \Psi^{\dagger}_{\k} h_\mathcal{B}(\k) \Psi_{\k}$, where 
\beq
 h_{\mathcal{B}}(\k) =  \begin{pmatrix}
\varepsilon(\k) & \xi(\k) \\
  \xi^{*}(\k) & \varepsilon(\k)
\end{pmatrix}, ~~~
 \varepsilon(\k) &=& T_{2x} \, \text{cos}(2k_x) + T_{2y} \, \text{cos}(2k_y) \nonumber \\
 \xi(\k) &=& -2T_{y} \, \text{sin}(k_y) + 4 i T_{x+y} \, \text{cos}(k_x) \text{cos}(k_y)
\eeq
The two bands are therefore given by 
\beq
\label{bosDisp}
E^{\pm}(\k) =  \varepsilon(\k) \pm |\xi(\k)| = T_{2x} \, \text{cos}(2k_x) + T_{2y}  \, \text{cos}(2k_y) \pm 2 \sqrt{T_y^2 \,\text{sin}^2(k_y) + 4 T_{x+y}^2 \text{cos}^2(k_x)\text{cos}^2(ky)} \nonumber \\
\eeq
In general, the minima of $E^{-}(\k)$, which corresponds to the momentum at which the boson condenses, will lie at some incommensurate point. In Fig. \ref{phaseDia}, we present an approximate phase diagram and look in more details into the different kinds of superconducting phases obtained by condensing the boson. All but one of these phases break time reversal symmetry $\T$. 

\subsubsection{$\mathcal{T}$-invariant PDW}
First, consider the case where we turn off the imaginary hopping terms, i.e, $T_y = T_{x+y} = 0$. In this case, the boson hoppings are translationally invariant, and the minima corresponds to $\Q = (0,0)$. Let the boson condensate at $\Q = (0, 0)$ be $\mathcal{B}(\r) = B_o$, we find that the nearest neighbor c-fermion pairing amplitude is given by 
\beq
 \Delta^c_{\r,\r+ \hat{x}}  &=& \mathcal{B}_o^2  (-1)^{y} \Delta_{1x} \nonumber \\
 \Delta^c_{\r,\r+ \hat{y}}  &=& \mathcal{B}_o^2 \, \Delta_{1y} 
\eeq
The superconducting phase breaks translation symmetry, therefore we have a PDW state with $\Q_{PDW} = (0,\pi)$.
Since the bosons condense at zero momentum, the density wave order parameter can only pick up a non-zero expectation value if the $f$ spinon hoppings themselves break translation symmetry. This is not the case for our fermionic ansatz [described by Eq.~(\ref{Fansatz1})], and therefore we expect no density wave order in this phase. In fact, one can perturbatively evaluate the renormalizations of the $c$ fermion hoppings (over and above the ones which are present in $H_c$) as follows: 
\beq
\langle c^{\dagger}_{\r} c_{\r + 2 \hat{x}} \rangle &=& \mathcal{B}_o^2 (-t_{2x}) \nonumber \\
\langle c^{\dagger}_{\r} c_{\r + 2 \hat{y}} \rangle &=& \mathcal{B}_o^2 (-t_{2y})
\eeq 
These are both translation invariant.

\subsubsection{Translationally invariant SC with broken $\mathcal{T}$}
$\Q = (0,0)$ is also the position of the minima when $T_y < T_{x+y}$. However, any non-zero $T_{x+y}$ will enlarge the unit cell. The value of the boson condensate is therefore given by 
\beq
\mathcal{B}(\r) =  \begin{pmatrix}
\mathcal{B}_A(\r) \\  \mathcal{B}_B(\r)
\end{pmatrix} = \mathcal{B}_o \begin{pmatrix}
1 \\  i
\end{pmatrix} 
\eeq

From the boson condensate at $\Q = (0, 0)$, we find that the nearest neighbor c-fermion pairing amplitude is given by 
\beq
 \Delta^c_{\r,\r+ \hat{x}}  &=& \mathcal{B}_o^2  (-1)^{y} \Delta_{1x}, \text{ for } \r \in A \nonumber \\
  \Delta^c_{\r,\r+ \hat{x}} &=&  (i \mathcal{B}_o)^2  (-1)^{y} \Delta_{1x} = - \mathcal{B}_o^2 (-1)^{y} \Delta_{1x}, \text{ for } \r \in B,  \text{ and } \nonumber \\
 \Delta^c_{\r,\r+ \hat{y}}  &=& i \mathcal{B}_o^2 \Delta_{1y} 
\eeq
Noting that there is the $A/B$ sublattices are defined by even/odd $y$ coordinates, this implies that $ \Delta^c_{\r,\r+ \hat{x}} = \mathcal{B}_o^2 \Delta_{1x}$. Thus, this superconductor does not break translation symmetry. However, it will break necessarily time-reversal symmetry because there is a relative $i$ between the pairing amplitudes along $\hat{x}$ and $\hat{y}$, and the pairing is of the $s + i d_{x^2 - y^2}$ type. This state does not have an associated density wave order. 

Depending on the relative signs of the hoppings, a condensate at $\Q = (\pi,0)$ is also possible, and gives a superconducting state with identical features. 

\subsubsection{Commensurate PDW with broken $\T$}
Next, let us consider the case where the nearest-neighbor hopping dominates, i.e, $T_y \gg T_{x+y}, T_{2x}, T_{2y}$. In this case, there is a regime where the minima of the boson dispersion lies approximately at $\pm \Q = (0, \pm \pi/2)$. The boson condensate is given by
\beq
\mathcal{B}(\r) =  \begin{pmatrix}
\mathcal{B}_A(\r) \\  \mathcal{B}_B(\r)
\end{pmatrix} = \mathcal{B}_+ \begin{pmatrix}
e^{i \Q \cdot \r_A } \\  e^{i \Q \cdot \r_B}
\end{pmatrix} + \mathcal{B}_- \begin{pmatrix}
e^{-i \Q \cdot \r_A } \\  e^{-i \Q \cdot \r_B}
\end{pmatrix} =  \mathcal{B}_+ \begin{pmatrix}
1 \\  i 
\end{pmatrix} e^{i \Q \cdot \r_A } +  \mathcal{B}_- \begin{pmatrix}
1 \\  -i 
\end{pmatrix} e^{-i \Q \cdot \r_A } 
\eeq
Using the previously outlined procedure to calculate the superconducting order parameter, we find 
\beq
\Delta^c_{\r,\r+ \hat{x}} &=& \left[ (\mathcal{B}_+^2 + \mathcal{B}_-^2) + (-1)^y 2 \mathcal{B}_+ \mathcal{B}_- \right] \Delta_{1x} \nonumber \\
\Delta^c_{\r,\r+ \hat{y}} & = &  i \left( \mathcal{B}_+^2 - \mathcal{B}_-^2 \right)  (-1)^y \Delta_{1y} 
\eeq
Both translation symmetry and time-reversal symmetry are explicitly broken by the superconductor, and we have a PDW at $\Q_{PDW} = (0,\pi)$ with $s + i d_{x^2 - y^2}$ pairing. 

Analogous to the first PDW phase with unbroken $\T$, we can evaluate the renormalization of the $c$ fermion hopping amplitudes (suppressing spin indices for simplicity):
\beq
\langle c^{\dagger}_{\r} c_{\r + 2 \hat{x}} \rangle &=& \left[ |\mathcal{B}_{+}|^2 + |\mathcal{B}_{-}|^2 +  (-1)^y  (\mathcal{B}_{+}\mathcal{B}_{-}^{*} + \mathcal{B}_{-}\mathcal{B}_{+}^{*} ) \right] (-t_{2x}) \nonumber \\
\langle c^{\dagger}_{\r} c_{\r + 2 \hat{y}} \rangle &=& \left[ (|\mathcal{B}_{+}|^2 + |\mathcal{B}_{-}|^2)(-1)^y + (\mathcal{B}_{+}\mathcal{B}_{-}^{*} + \mathcal{B}_{-}\mathcal{B}_{+}^{*} ) \right] (-t_{2y})
\eeq
The spatially constant parts of the induced hopping amplitudes will just renormalize the bare hopping of the $c$ fermions, but the terms at $\Q_{CDW} = \Q_{PDW} = (0,\pi)$ correspond to a density wave with form factor $P_{\Q_{CDW}}(\k) = c_1 \, \text{cos}(2k_x) + c_2 \, \text{cos}(2k_y)$, which is of the $s^{\prime} + d$ type. This is therefore an example of a state where PDW co-exists with bond density wave order. 

\subsubsection{Incommensurate PDW with broken $\T$}
Away from the previous two parameter regimes, the boson $b(\r)$ will condense at some generic incommensurate momentum $\Q = (Q_x, Q_y)$. One can carry out an analogous calculation to find out the relevant order parameters. Note that the boson dispersion is symmetric under $\k \rightarrow -\k$, which implies that there are necessarily a couple of minima at $\Q$ and $-\Q$. Assuming no other degenerate minima, the boson condensate is given by:
\beq
\mathcal{B}(\r) =  \begin{pmatrix}
\mathcal{B}_{A+} e^{i\Q \cdot \r_A} \\  \mathcal{B}_{B+} e^{i\Q \cdot \r_B}
\end{pmatrix}  + \begin{pmatrix}
\mathcal{B}_{A-} e^{i\Q \cdot \r_A} \\  \mathcal{B}_{B-} e^{i\Q \cdot \r_B}
\end{pmatrix} 
\eeq
This leads to a PDW at momentum  $2\Q + (0,\pi)$ as well as $(0,\pi)$ for the $c$-fermions, the latter coming from the inherent translation symmetry breaking of the spinon pairing ansatz:
\beq
\Delta^c_{\r,\r+ \hat{x}} &=& \left[ \mathcal{B}_{A+}^2 \, e^{i ( 2\Q\cdot\r + Q_x)} + 4 \mathcal{B}_{A+} \mathcal{B}_{A-} \text{cos}(Q_x) +  \mathcal{B}_{A-}^2 \, e^{-i ( 2\Q\cdot\r + Q_x)} \right] (-1)^y \Delta_{1x} , \r \in A \nonumber \\
&=& \left[ \mathcal{B}_{B+}^2 \, e^{i ( 2\Q\cdot\r + Q_x)} + 4 \mathcal{B}_{B+} \mathcal{B}_{B-} \text{cos}(Q_x) +  \mathcal{B}_{B-}^2 \, e^{-i ( 2\Q\cdot\r + Q_x)} \right] (-1)^y \Delta_{1x} , \r \in B \nonumber \\
\Delta^c_{\r,\r+ \hat{y}} & = & \left[ \mathcal{B}_{A+}  \mathcal{B}_{B+} \, e^{i (2\Q\cdot\r + Q_y)} +  \mathcal{B}_{A-}  \mathcal{B}_{B+} \, e^{i Q_y} +  \mathcal{B}_{A+}  \mathcal{B}_{B-} \, e^{-iQ_y} +  \mathcal{B}_{A-}  \mathcal{B}_{B-} \,  e^{-i (2\Q\cdot\r + Q_y)} \right] \Delta_{1y} \nonumber \\
\eeq
An analogous calculation of the density wave order parameter shows that there is an oscillation of charge density on the bonds at momenta $\Q_{CDW} = 2\Q$. 
\beq
\langle c^{\dagger}_{\r} c_{\r + 2 \hat{x}} \rangle &\sim& \mathcal{B}_{A/B}^2 e^{2 i \Q \cdot \r} (-t_{2x}) , ~~ \r \in A/B  \nonumber \\
\langle c^{\dagger}_{\r} c_{\r + 2 \hat{y}} \rangle &\sim& \mathcal{B}_{A/B}^2  e^{2 i \Q \cdot \r} (-t_{2y}) , ~~ \r \in A/B 
\eeq
Therefore, we have an incommensurate PDW co-existing with bond density wave.

More generally, boson condensation at two different momenta $\Q$ and $\Q^{\prime}$ will lead to a PDW order at $\K_{PDW} = \Q + \Q^{\prime} + (0,\pi)$ and $(0,\pi)$, and a bond density wave order at momenta $\K_{CDW} = \Q \pm \Q^{\prime}$ for our fermionic ansatz. These are all states with co-existing PDW and density wave order. Note that a density wave at a different momentum $\Q_{DW} = \Q + \Q^{\prime} + \K_1$ is also possible if there is a spinon-hopping term which breaks translation symmetry with momentum $\K_1$. In our fermionic ansatz for the $f$ spin liquid, such a term is absent (upto NNNN) and therefore such a density wave does not exist.

We now argue that inclusion of $T^{od}_{\alpha}$ does not change these phases, although it enlarges the phase space and therefore can change where these show up in the phase space. This can be explicitly seen from the eigenvalues of the $4\times4$ matrix $h(\k)$ in momentum space, which are now given by (assuming $T^{d/od}_{2x} = T^{d/od}_{2y} = T^{d/od}_{2}$ to avoid clutter of notation):
\beq
E^{+}_{\k,\pm} = 2[ \text{cos}(2k_x) + \text{cos}(2k_y)](T^{d}_{2} - T^{od}_2) \pm 2\sqrt{(T^{d}_{y} - T^{od}_y)^2 \text{sin}^2(k_y) + 4 (T^{d}_{x+y} - T^{od}_{x+y})^2 \text{cos}^2(k_x)\text{cos}^2(k_y)}  \nonumber \\
E^{-}_{\k, \pm} = 2[ \text{cos}(2k_x) + \text{cos}(2k_y)](T^{d}_{2} + T^{od}_2) \pm 2\sqrt{(T^{d}_{y} + T^{od}_y)^2 \text{sin}^2(k_y) + 4 (T^{d}_{x+y} + T^{od}_{x+y})^2 \text{cos}^2(k_x)\text{cos}^2(k_y)} \nonumber \\
\eeq
These are essentially identical to the previous dispersion in Eq.~(\ref{bosDisp}), with a renormalization of hopping parameters. Therefore, condensates again occur at the same values of $\Q$ as described previously, and lead to the same phases.


\section{Conclusions}
\label{sec:conc}

While several recent experiments \cite{Marel13,MG14} have been consistent with a FL* model for the pseudogap metal at higher temperatures, the most
recent Hall effect measurements \cite{LTCP15}
indicate that the FL* model may well extend down to low temperatures just below optimal doping.

In the light of this, it is useful to catalog the confinement instabilities of the simplest FL* state, the $\mathbb{Z}_2$-FL*. 
The excitations of this state invariably transform non-trivially under global symmetries of the model, and so the confinement transition
is then {\it simultaneous\/} with some pattern of symmetry breaking. From Table~\ref{tab:z2}, we observe that the 
$\mathbb{Z}_2$-FL* state has three categories of bosonic excitations, and each can then give rise to a distinct confinement transition.
The most familiar is the condensation of the bosonic spinons (column $e$ in Table~\ref{tab:z2}), and this leads to spin-density-wave order,
which is observed in most cuprates at low doping. The second possibility is the condensation of visons (column $m$ in Table~\ref{tab:z2}):
this was examined recently \cite{PCAS16}, and it was found that bond-density-waves similar to recent 
observations \cite{Fujita14,Comin14sym,Forgan15}
are a possible outcome. The final class of confinement transitions out the $\mathbb{Z}_2$-FL* state was considered in the present paper:
this is the condensation of bosonic chargons (column $\epsilon_c$ in Table~\ref{tab:z2}). 

Our main technical challenge in this paper was to compute the projective symmetry group of the fermionic spinons (column $\epsilon$ in Table~\ref{tab:z2}) for a favorable $\mathbb{Z}_2$ spin liquid state described by an ansatz for bosonic spinons \cite{ReadSachdev,ReadSachdev2,SubirQPT}. 
An important feature of the PSG for the fermionic spinons obtained was that translational symmetry was realized projectively,
with $T_x T_y = - T_y T_x$. 
After obtaining this PSG, we could then deduce the PSG for the bosonic chargons
by fusing the fermionic spinons to the electron, which has a trivial PSG. The 
PSG for the bosonic chargons also had $T_x T_y = - T_y T_x$, and this almost always means that the confinement state with
condensed chargons will break translational symmetry. Combined with the pairing of fermionic spinons invariably present in the
$\mathbb{Z}_2$-FL* state, such analyses led to the appearance of FFLO, or pair density wave (PDW), superconductivity. 
And it is worthwhile to note here the recent observation of modulated superconductivity, albeit on a much larger background
of uniform superconductivity \cite{Hamidian16}.

In conclusion, we highlight the remarkable fact that the three categories of confinement transitions out of $\mathbb{Z}_2$-FL*
allowed by Table~\ref{tab:z2} (corresponding to the three columns with bosonic self-statistics) correspond closely to 
features of the phase diagrams of the cuprates: ({\it i\/}) the condensation of $m$ can lead
to metals with density wave order similar to observations, as discussed recently in Ref.~\onlinecite{PCAS16};
({\it ii\/}) the condensation of $e$ leads to incommensurate magnetic order found at low doping;
({\it iii\/}) the present paper showed show the condensation of $\epsilon_c$ can lead to superconductors
with co-existing density wave order, a state observed in recent experiments \cite{Hamidian16}.

\section*{Acknowledgments} 
We acknowledge helpful conversations with Y. M. Lu and F. Wang. 
This research was supported by the NSF under Grant DMR-1360789. J. S. was supported by the National Science Foundation Graduate Research Fellowship under Grant No. DGE1144152.
Research at Perimeter Institute is supported by the Government of Canada through Industry Canada 
and by the Province of Ontario through the Ministry of Research and Innovation.   

\appendix
\section{Derivation of the bosonic PSG}
\label{bosPSG}
To derive the solution, we note a few things. First, if we apply a gauge transformation $G$ to the ansatz, then the gauge transformed ansatz is invariant under 
\begin{equation}
GG_{X}XG^{-1}=GG_{X}XG^{-1}X^{-1}X \implies G_X \rightarrow GG_{X}XG^{-1}X^{-1}
\end{equation}
This implies that the phase $\phi_X$ under a gauge transformation transforms as (except when $X$ is the anti-unitary time-reversal operator)
\begin{equation}
\label{gaugeTransPhase}
\phi_{X}(\textbf{r}) \rightarrow \phi_{G}(\textbf{r})+\phi_{X}(\textbf{r})-\phi_{G}[X^{-1}(\textbf{r})]
\end{equation}
Since we can choose a particular gauge to work in, we shall use this to later simplify our PSG classification.

Let us find the constraints imposed by the structure of the rectangular lattice symmetry group. Consider a string of space group operators which combine to identity in the lattice symmetry group. Then in the PSG, these must combine to an element of the IGG $\mathbb{Z}_2$, which means it is $\pm 1$. Therefore, for each such string, we shall define an integer $p_n$ (defined modulo 2) which will denote how the symmetry fractionalizes in the PSG. It is sufficient to consider the strings in Eqs.~(\ref{crysSymend}), because any other string can be reduced to one such string by \textit{normal ordering} the strings using the same commutation/anticommutation relations. We can then use these constraints to find the gauge operations $G_X$, or equivalently, their phases $\phi_X(\bf{r})$, in terms of the $p_n$'s. Note that all the following equations for the phases are true modulo $2\pi$. For notational convenience, we also introduce discrete lattice derivatives $\Delta_{x} \phi_{X} = \phi_X(x+1,y) - \phi_X(x,y)$, and $\Delta_{y} \phi_{X} = \phi_X(x,y+1) - \phi_X(x,y)$.

Let us start by looking at the commutation relation between the translations. We have, from Eq.~(\ref{crysSymbegin})
 \begin{equation}
 (G_{T_x}T_{x} )^{-1} (G_{T_y}T_{y})(G_{T_x}T_{x})(G_{T_Y}T_{y})^{-1}=(T^{-1}_{x}G_{T_{x}}T_{x})(T^{-1}_{x}G_{T_y}T_{x})(T^{-1}_{x}T_{y}G_{T_{x}}T^{-1}_{y}T_{x})(G^{-1}_{Ty})=\pm 1 = (-1)^{p_1}
 \end{equation}
Since $Y^{-1}G_{X}Y: \phi_{X}(\textbf{r}) \rightarrow \phi_{X}[Y(\textbf{r})]$, we have the following constraint equation for $\phi_{T_{x}}$ and $\phi_{T_{y}}$
\begin{equation}
-\phi_{Tx}\left[T_{x}(x,y)\right]+\phi_{Ty}\left[T_{x}(x,y)\right]+\phi_{Tx}\left[T^{-1}_{y}T_{x}(x,y)\right]-\phi_{T_{y}}(x,y)=p_{1}\pi
\end{equation}
Now we assume we are defining the system on open boundary conditions, so that we can use the gauge freedom in Eq.~(\ref{gaugeTransPhase}) to set $\phi_{T_{x}}(x,y)=0$. We also assume, following Ref.~\onlinecite{WangVishwanath} that we can set $\phi_{T_{y}}(0,y)=0$. Then we can write down the solution as
\begin{equation}
\Delta_{x} \phi_{T_y}(x,y) = p_1 \pi \implies \phi_{T_y}(x,y) = p_1 \pi x + \phi_{T_y}(0,y) = p_1 \pi x
\end{equation}
Now we consider $P_x$ and its commutations with $T_x$ and $T_y$. From $G_{T_x} T_x P_x^{-1} G_{P_x}^{-1} G_{T_x} T_x G_{P_x} P_x = \pm 1 = (-1)^{p_2}$, we get
\begin{eqnarray}
\phi_{P_x}(x,y) - \phi_{P_x}[T_xP_x(x,y)] + \phi_{T_x}[P_x(x,y)] + \phi_{P_x}[P_x(x,y)] = p_2 \pi \nonumber 
\implies \Delta_x \phi_{P_x} = p_2 \pi
\end{eqnarray}
From $G_{T_y}^{-1} T_y P_x^{-1} G_{P_x}^{-1} G_{T_y} T_y G_{P_x} P_x = \pm 1 = (-1)^{p_4}$, we get
\begin{eqnarray}
- \phi_{T_y}[T_y(x,y)] - \phi_{P_x}[P_x T_y(x,y)] + \phi_{T_y}[T_y P_x(x,y)] + \phi_{P_x}[P_x(x,y)] = p_4 \pi \nonumber \\
\implies \Delta_y \phi_{P_x} - p_1 \pi (-x) + p_1 \pi (-x) = p_4 \pi
\implies \Delta_y \phi_{P_x}  = p_4 \pi
\end{eqnarray}
Using the above two equations, we can write down
\begin{equation}
\phi_{P_x}(x,y) = p_2 \pi x + p_4 \pi y + \phi_{P_x}(0,0)
\end{equation} 
$\phi_{P_x}(0,0)$ is now found out using $(G_{P_x}P_x)^2 = \pm 1 = (-1)^{p_6}$, which implies $2\phi_{P_x}(0,0) = p_6 \pi$
\begin{equation}
\phi_{P_x}(x,y) = p_2 \pi x + p_4 \pi y + \frac{p_6}{2} \pi 
\end{equation}
In an exactly analogous way, we find that 
\begin{equation}
\phi_{P_y}(x,y) = p_3 \pi x + p_5 \pi y + \frac{p_7}{2} \pi 
\end{equation}
Finally, let us consider time-reversal $\mathcal{T}$. From the commutations of $\mathcal{T}$ with $T_x$ and $T_y$, we find the following two equations
\begin{equation}
\Delta_x \phi_{\mathcal{T}} = p_8 \pi, \; \Delta_y \phi_{\mathcal{T}} = p_9 \pi
\end{equation}
Solving the above gives us $\phi_{\mathcal{T}}(x,y) = p_8 \pi x + p_9 \pi y + \phi_{\mathcal{T}}(0,0)$. 
The commutations with $P_x$ and $P_y$ do not yield any new relation. Finally, we note that under a global gauge transformation $G: b_{\r \sigma}  \rightarrow e^{i\theta} b_{\r \sigma}$, due to the anti-unitary nature of $\mathcal{T}$, we have $ \phi_{\mathcal{T}}(x,y) \rightarrow \phi_{\mathcal{T}}(x,y) + 2\theta$. We can use this freedom to set $\theta = - \phi_{\mathcal{T}}(0,0)/2$, and we therefore have 
\begin{equation}
\phi_{\mathcal{T}}(x,y)  =  p_8 \pi x + p_9 \pi y 
\end{equation}
Note that this gauge transformation does not affect the $\phi_X$ corresponding to a spatial symmetry $X$, as these are unitary and follow Eq.~(\ref{gaugeTransPhase}).

\section{PSG corresponding to the nematic bosonic ansatz}
\label{BosAnsPSG}
\begin{figure}[h]
\centering
{
\subfigure[]{\includegraphics[width=0.27\textwidth]{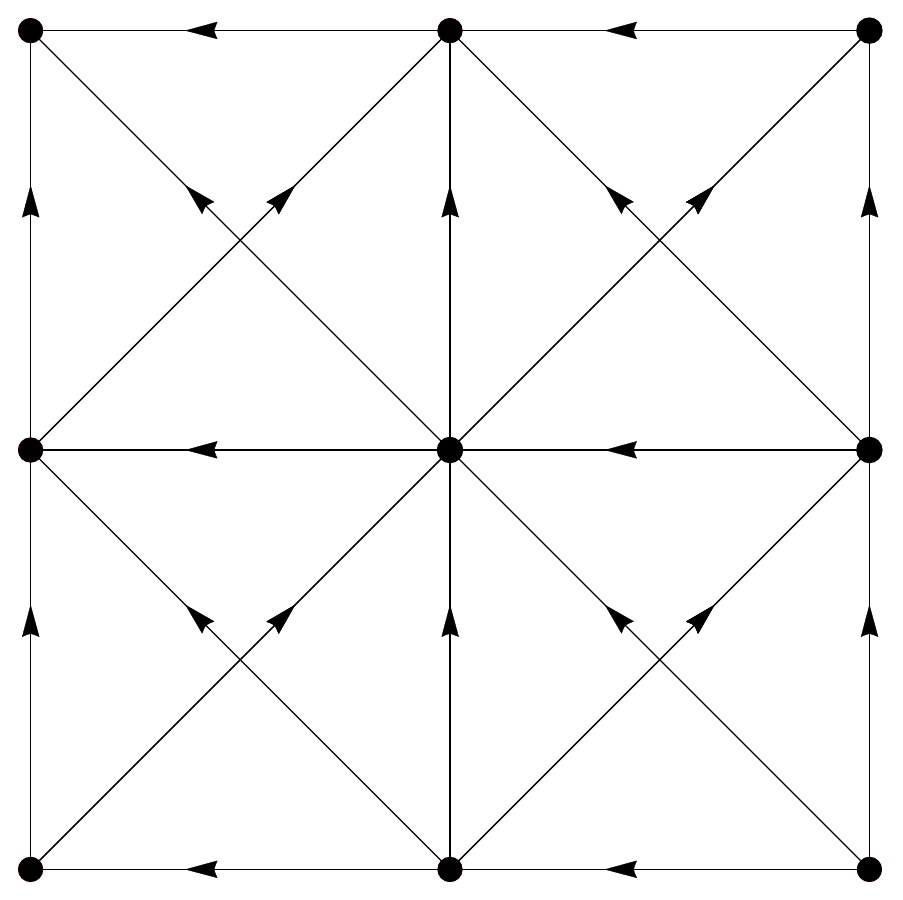} \label{fig:ansatz}} \subfigure[]{\includegraphics[width=0.27\textwidth]{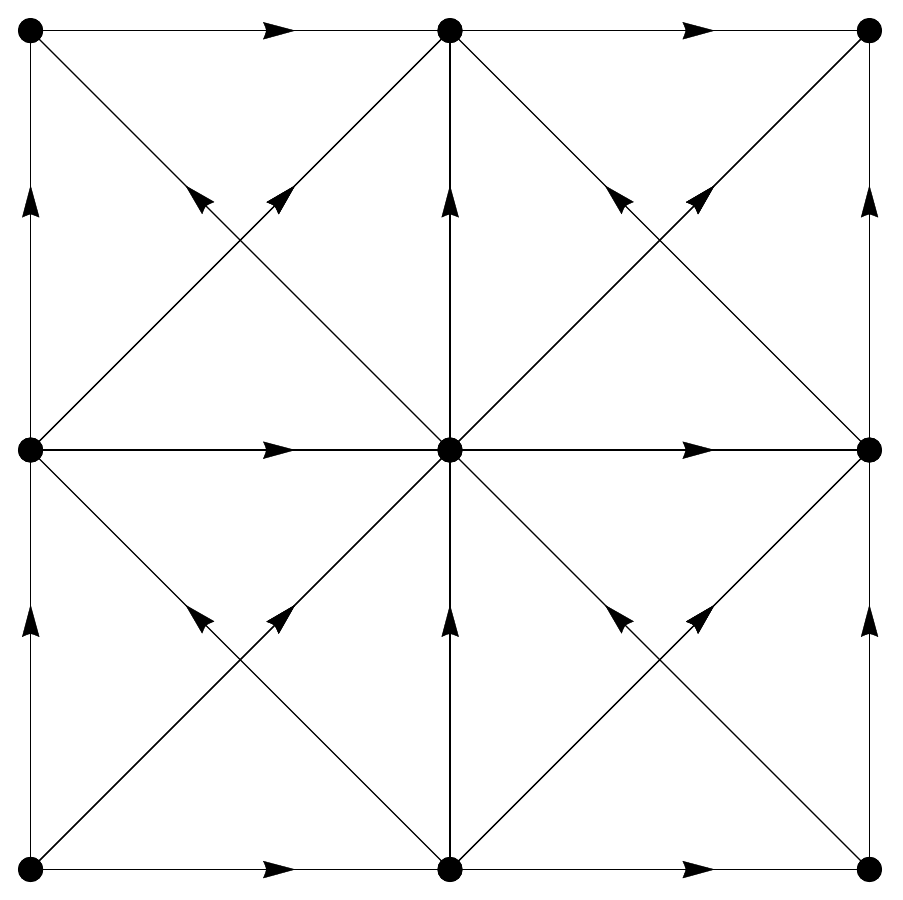} 
\label{fig:px}}
\subfigure[]{\includegraphics[width=0.27\textwidth]{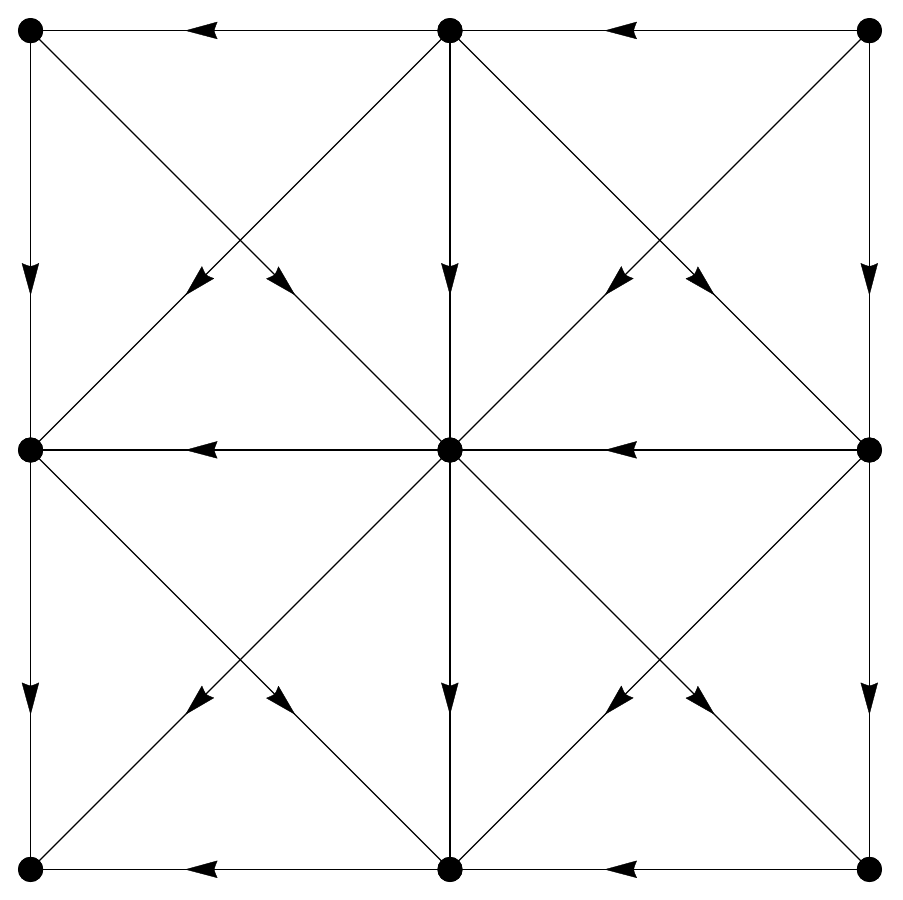} \label{fig:py}}
}
\caption{\subref{fig:ansatz} The original translation invariant ansatz \subref{fig:px} the ansatz under $P_{x}:(x,y) \rightarrow (-x,y)$ \subref{fig:py} the ansatz under $P_{y}:(x,y) \rightarrow (x,-y)$. The arrow from $\r$ to $\rp$ indicates the orientation for which $Q_{\r \rp} >0$. }
\end{figure}

The phases $\phi_X$ corresponding to the symmetry operations $X$ can be fixed by demanding that the ansatz remain invariant under $G_X X$. First, we note that the ansatz itself is translation invariant (see Fig.~\ref{fig:ansatz}), so both $G_{T_x}$ and $G_{T_y}$ must be trivial. This implies that our ansatz is consistent with our trivial gauge choice for $G_{T_x}$, and $p_1 =0$. 

\begingroup
\renewcommand*{\arraystretch}{2}
\begin{table}[ht]
\begin{tabular}{|c|c |}
\hline
$P_{x}$ & $P_{y}$
\\
\hline
$Q_{(x, y)\rightarrow (x+1, y)}$ $\rightarrow$ $Q_{(x+1, y)\rightarrow (x, y)} = -Q_{(x, y)\rightarrow (x+1, y)}$ & $Q_{(x, y) \rightarrow (x+1, y)}$ $\rightarrow$ $Q_{(x, y+1) \rightarrow (x+1, y+1)} =  Q_{(x, y) \rightarrow (x+1, y)}$
\\
$Q_{(x, y)\rightarrow (x, y+1)}$ $\rightarrow$ $Q_{(x+1, y)\rightarrow (x+1, y+1)} = Q_{(x, y)\rightarrow (x, y+1)}$ & $Q_{(x, y)\rightarrow (x, y+1)}$ $\rightarrow$ $Q_{(x, y+1)\rightarrow (x, y)} = -Q_{(x, y)\rightarrow (x, y+1)}$
\\
$Q_{(x, y)\rightarrow (x+1, y+1)}$ $\rightarrow$ $Q_{(x+1, y)\rightarrow (x, y+1)}=Q_{(x, y)\rightarrow (x+1, y+1)}$ & $Q_{(x, y)\rightarrow (x+1, y+1)}$ $\rightarrow$ $Q_{(x, y+1)\rightarrow (x+1, y)} = -Q_{(x, y)\rightarrow (x+1, y+1)}$
\\
$Q_{(x+1, y)\rightarrow (x, y+1)}$ $\rightarrow$ $Q_{(x, y)\rightarrow (x+1, y+1)} = Q_{(x+1, y)\rightarrow (x, y+1)}$
& $Q_{(x+1, y)\rightarrow (x, y+1)}$ $\rightarrow$ $Q_{(x+1, y+1)\rightarrow (x, y)} = -Q_{(x+1, y)\rightarrow (x, y+1)}$
\\
\hline
\end{tabular}
\caption{Transformation of link variables $Q_{\r \rp}$}
\label{table:links}
\end{table}
\endgroup

Let us now consider $P_x$. Using translation invariance, we have $P_x(Q_{\r,\r+\hat{x}}) = Q_{\r+\hat{x},\r} = -Q_{\r,\r+\hat{x}}$. By definition, $G_{P_x}P_x(Q_{\r,\r+\hat{x}}) = Q_{\r,\r+\hat{x}}$, and this implies that $\phi_{P_x}[P_x(\r)] + \phi_{P_x}[P_x(\r + \hat{x})] = \pi$, which in turn gives us $p_2 + p_6 = 1$. The nearest-neighbor $y$ bond is unaffected by $P_x$, whereas the diagonal bonds are swapped and effectively not affected as they have the same value in this ansatz. We get the following equations from demanding that $G_X$ acts trivially on these bonds: $p_4 + p_6 =0$, and $p_2 + p_4 + p_6 = 0$. Solving these we find that $p_2 =0, \, p_4 = p_6 = 1$ (modulo 2).
 
Similarly, acting $P_y$ changes the sign on all bonds except the $x$ bonds, and we have the following equations: $p_3 + p_7 = 0, \, p_5 + p_7 =1, \text{ and } p_3 + p_5 + p_7 =1$. Solving gives us $p_3 = p_7 = 0, \, p_5 = 1$. The transformations of the Ansatz under reflections are schematically described in Table \ref{table:links}.

Finally, we look at time-reversal. Since all the bond variables are real (which we assume is consistent with our gauge choice), we have $p_8 = p_9 =0$. 

\section{Alternate derivation of the vison PSG}
\label{visPSG}
In this section, we present an alternate derivation of the vison PSG, based on the critical modes of the vison as one approaches vison condensation. We assume a soft spin formulation, which is reasonable from coarse graining near a critical point. We replace the Ising variables $\tau^{z}_{\R}$s in the vison Hamiltonian by real fields $\phi_{\R} \in \mathbb{R}$, and describe the kinetic term by a conjugate momentum $\pi_{\R}$ to $\phi_{\R}$ and mass $m$, so that the Hamiltonian becomes
\begin{equation}
H_{soft} = \frac{1}{2}\sum_{\R} \left( \pi_{\R}^2 + m^2 \phi_{\R}^2 \right) + \sum_{\R \Rp} J_{\R \Rp} \, \phi_{\R} \,\phi_{\Rp}
\end{equation}

In our gauge choice (recall Fig. \ref{FFIMgauge}), we have a two-site unit cell with primitive vectors $\mathbf{a}_1 = \hat{x} + \hat{y}$ and $\mathbf{a}_2 = 2 \hat{y}$ (setting lattice spacings = 1). Neglecting the kinetic term (which is inessential to the study of  vison condensation transitions), the Hamiltonian in the momentum space for this extended unit cell is given by 
\begin{equation}
H_{soft} = \sum_{\mathbf{k}} H(\mathbf{k}), \text{ with } H(\mathbf{k}) = 2
\begin{pmatrix}
0 & \text{cos}k_y + i \text{ sin} k_x \\
\text{cos}k_y - i \text{ sin}k_x & 0
\end{pmatrix}
\end{equation}
Diagonalizing this leads to the following two bands
\begin{equation}
\omega_{\pm}(\mathbf{k}) = \pm 2 \sqrt{ \text{cos}^2k_y + \text{ sin}^2k_x}
\end{equation}
The inequivalent minima of this band structure lie at $\Q_{1,2} = \pm(\pi/2,0)$ in the reduced BZ, and the corresponding eigenvectors are $\mathbf{v}^{1} = (- e^{i \pi/4},1)^{T}$ and $\mathbf{v}^{2} = (- e^{-i \pi/4},1)^{T}$, where the superscript $T$ indicates transposition. Later, we shall write out the vison field in terms of these soft modes. 

Now, we analyze the PSG of the visons. Since the Hamiltonian is invariant under symmetry transformations only up to a gauge transformation, we identify, for each symmetry generator $X$ in the space group of the rectangular lattice, an element $G_X \in \mathbb{Z}_2$ such that $J_{\r \rp} = J_{X[\r]X[\rp]} G_X[X(\r)]G_X[X(\rp)]$. 
These symmetry operations for the rectangular lattice, and their associated gauge transformations are listed below. We denote sublattice $s =(1,2)$ at the unit cell $\mathbf{r} = m \, \mathbf{a}_1 + n \, \mathbf{a}_2$ by $(m,n)_s$.
\begin{eqnarray}
T_x:  \begin{cases}
(m,n)_1 \rightarrow (m+1,n-1)_2 \\
(m,n)_2 \rightarrow (m+1,n)_1
\end{cases} \nonumber \\
T_y:  \begin{cases}
(m,n)_1 \rightarrow (m,n)_2 \\
(m,n)_2 \rightarrow (m,n+1)_1
\end{cases} \nonumber \\
P_x:  \begin{cases}
(m,n)_1 \rightarrow (-m-1,m+n)_2 \\
(m,n)_2 \rightarrow (-m-1,m+n+1)_1
\end{cases} \nonumber \\
P_y:  \begin{cases}
(m,n)_1 \rightarrow (m,-n-1)_2 \\
(m,n)_2 \rightarrow (m,-n-1)_1
\end{cases} 
\end{eqnarray}
The associated gauge transformations can be found out by figuring out appropriate gauge transformations to leave the Hamiltonian invariant. As discussed in the main text, all operations except $P_x$ exchange the $x$ bonds with different signs, and hence need a gauge transformation which adds an extra sign to bring the Hamiltonian back to itself. The $y$ bonds are invariant under any of these operations. 
\begin{eqnarray}
G_{T_x}(m,n)_s &=& (-1)^{m} \nonumber \\
G_{T_y}(m,n)_s &=& (-1)^{m} \nonumber \\
G_{P_x}(m,n)_s &=& 1 \nonumber \\
G_{P_y}(m,n)_s &=& (-1)^{m}
\end{eqnarray}
Next, we outline to find the general procedure to find the representation of the PSG in the order parameter space, and subsequently apply it to our situation. We first define the order parameter by expanding the vison field in terms of the $N$ soft modes as follows:
\begin{equation}
\phi_{s}(\mathbf{R}) = \sum_{n=1}^{N} \psi_{n} v^{n}_{s} e^{i \mathbf{q}_n\cdot\mathbf{R}}
\end{equation}
Here, $\mathbf{R}$ is the unit cell index, $s = (1,2)$ is the sub lattice index, $N$ is the number of soft modes and the complex number $\psi_n$ is the vison order parameter corresponding to the $n$th soft mode at momentum $\mathbf{q}_n$ with eigenvector $\mathbf{v}^{n}$ of $H_{soft}$. Now, we can figure out how the order parameters $\psi_n$ transform into each other under different symmetry operations $G_X X$ which leave the Hamiltonian $H_{soft}$ invariant. This can be found from solving the following equation, which gives us the desired representation in form of the $N \times N$ matrix $O_{X}$ defined below [with $(\Rp,s^{\prime}) = X(\r,s)$]:
\begin{eqnarray}
G_X X[\phi_s(\R)] &=& \sum_{n=1}^{N} \psi_n v_{s^{\prime}}^{n} e^{i \mathbf{q}_n\cdot\Rp} G_X[\mathbf{R}^{\prime},s^{\prime}] \nonumber \\
& = & \sum_{n=1}^{N} \psi_n^{\prime} v^{n}_{s}  e^{i \mathbf{q}_n\cdot\R} \nonumber \\
& = & \sum_{n=1}^{N} \left( \sum_{m=1}^{N} O_{X,mn} \psi_n \right) v^{n}_{s}  e^{i \mathbf{q}_n\cdot\R}.
\end{eqnarray}
With nearest neighbor interactions of the soft spins in the fully frustrated dual Ising model, we earlier found that there are two minima at $\mathbf{Q}_{1,2} = \pm \mathbf{Q} = (\pm \pi/2,0)$ with associated eigenvectors $\mathbf{v}^{1}$ and $\mathbf{v}^2$. Since the order parameter $\phi$ is real, we can write it (in form of a vector with two sub lattice indices)
\begin{equation}
\begin{pmatrix}
\phi_1 \\
\phi_2
\end{pmatrix} = \psi \begin{pmatrix}
-e^{i \pi/4} \\ 1
\end{pmatrix} e^{i \mathbf{Q} \cdot \R} + \psi^{*} \begin{pmatrix}
-e^{-i \pi/4} \\ 1
\end{pmatrix} e^{-i \mathbf{Q} \cdot \R}.
\end{equation}
We work out the results for $T_x$ explicitly, and just quote the other ones. All of these can be obtained by following the general procedure outlined above. For $\mathbf{r} = (m,n)$, we have $\mathbf{Q} \cdot \R = \pi m/2$, so we get 
\begin{eqnarray}
\phi_1(\mathbf{R}) &=& - \psi e^{i \pi/4} e^{i \pi m/2} - \psi^{*} e^{-i \pi/4} e^{- i \pi m/2} \nonumber \\
\implies G_{T_x}T_x [\phi_1(\mathbf{R})]  &=& \left[ \psi (1) e^{i \pi/2 (m+1)} +  \psi^{*} (1) e^{- i \pi /2(m + 1)} \right] (-1)^m \nonumber \\
&=& \psi \, e^{i \pi/2} e^{-i \pi m/2} + \psi^{*} \, e^{-i \pi/2} e^{ i \pi m/2}  \nonumber \\
& = & - \psi^{\prime} e^{i \pi/4} e^{i \pi m/2} - \psi^{\prime *} e^{-i \pi/4} e^{- i \pi m/2}.
\end{eqnarray}
Since the above is true for all $m$, we have $\psi^{\prime} = - \psi^{*} e^{-i 3\pi/4} = e^{i \pi/4} \psi^{*}$. Therefore, in the matrix form, we can write 
\begin{equation}
\begin{pmatrix}
\psi^{\prime} \\
\psi^{\prime *}
\end{pmatrix}  = \begin{pmatrix}
0 & e^{i \pi/4} \\
e^{-i \pi/4} & 0 \\
\end{pmatrix} \begin{pmatrix}
\psi \\
\psi^{*}
\end{pmatrix}.
\end{equation}
Thus the matrix representation of $O_{T_x}$ in the order parameter space (in our chosen gauge) is given by 
\begin{equation}
O_{T_x} =  \begin{pmatrix}
0 & e^{i \pi/4} \\
e^{-i \pi/4} & 0 \\
\end{pmatrix} .
\end{equation}
The matrix representations of the other operators are worked out identically, here we just list the results. 
\begin{eqnarray}
O_{T_y} & = & \begin{pmatrix}
0 & -e^{-i \pi/4} \\
-e^{i \pi/4} & 0 \\
\end{pmatrix}, \\ 
O_{P_x} & = & \begin{pmatrix}
0 & e^{i \pi/4} \\
e^{-i \pi/4} & 0 \\
\end{pmatrix}, \\
O_{P_y} & = & \begin{pmatrix}
0 & -e^{-i \pi/4} \\
-e^{i \pi/4} & 0 \\
\end{pmatrix} . 
\end{eqnarray}
The fractionalization of the commutation relations can now be obtained from these matrices.
\begin{subequations}
\begin{eqnarray}
O_{T_x}O_{T_y}O_{T_x}^{-1}O_{T_y}^{-1} = -1, \\
O_{T_x}O_{P_x}O_{T_x}O_{P_x}^{-1} = 1, \\
O_{T_x}O_{P_y}O_{T_x}^{-1} O_{P_y}^{-1} = - 1, \\
O_{T_y}O_{P_x} O_{T_y}^{-1}O_{P_x}^{-1} = -1, \\
O_{T_y}O_{P_y}O_{T_y} O_{P_y}^{-1} = 1, \\
O_{P_x} O_{P_x} = 1, \\
O_{P_y} O_{P_y} = 1, \\
O_{P_x}O_{P_y}O_{P_x}^{-1}O_{P_y}^{-1} = -1.
\end{eqnarray}
\end{subequations}

A more complicated analysis including fourth-nearest-neighbor interactions \cite{WangSqLattice} (done on the square lattice, but works for rectangular lattices as well) also leads to matrix representations of the operators with identical crystal symmetry fractionalization. 

In order to check how the symmetries involving time-reversal fractionalize, we follow Ref. \onlinecite{Unification_Luetal}. We look at the edge modes and require that they are not symmetry protected, or, in other words, we have a gapped boundary. The edge modes of a $\mathbb{Z}_2$ spin liquid can always be fermionized with the same number of right and left movers (branch denoted by index $n$),
\begin{equation}
\mathcal{L}_{edge,0} = \sum_{n} i \psi_{L,n}^{\dagger} (\partial_t - v \partial_x) \psi_{L,n} -  i \psi_{R,n}^{\dagger} (\partial_t + v \partial_x) \psi_{R,n} .
\end{equation}
In general, we would expect a gapped edge due to backscattering terms below, unless these are forbidden by symmetry.
\begin{equation}
\mathcal{L}_{edge,1} = \sum_{m,n} \psi_{L,m}^{\dagger} M_{m,n} \psi_{R,n} +  \psi_{L,m}^{\dagger} \Delta_{m,n}\psi_{R,n}  + \mbox{H.c}
\end{equation}
The above mass terms correspond to condensing spinons or visons at the edge. Since condensing spin-half spinons would break SU(2) symmetry, we would need to condense visons to get gapped edges with all symmetries intact. This can only take place if the vison PSGs allow a vison condensate at the edge. If the symmetries act non-trivially on the vison field $\phi$, then the vison condensate will break the symmetry. Therefore, if we want to preserve the symmetry at the edge with gapped edge modes (non-zero mass terms), the symmetries at the edge cannot have a non-trivial action on $\phi$.

Consider the square lattice on a cylinder with open boundaries parallel to $\hat{x}$. Then the remaining symmetries are $T_x$, $P_x$ and time-reversal $\mathcal{T}$. If there are no symmetry-protected gapless edge states on the boundary, then these symmetries must act trivially on the visons. Hence, we have 
\begin{equation}
O_{T_x}^{-1} O_{\mathcal{T}}^{-1} O_{T_x} O_{\mathcal{T}} = 1, \; O_{P_x}^{-1} O_{\mathcal{T}}^{-1} O_{P_x} O_{\mathcal{T}} = 1
\end{equation}
We can apply an analogous argument for a cylinder with open boundaries parallel to $\hat{y}$, to find
\begin{equation}
O_{T_y}^{-1} O_{\mathcal{T}}^{-1} O_{T_y} O_{\mathcal{T}} = 1, \; O_{P_y}^{-1} O_{\mathcal{T}}^{-1} O_{P_y} O_{\mathcal{T}} = 1
\end{equation}

\section{Derivation of the fermionic PSG}
\label{fermPSG}
To derive the general solutions to the fermionic PSG, we note that the PSGs of two gauge-transformed ansatz are related (similar to the bosonic case). Recall that the PSG is defined as the set of all transformations $G_X X$ that leave the ansatz unchanged.
\begin{equation}
G_X X (U_{\r \rp}) = G_X \left( U_{X[\r]X[\rp]} \right) = U_{\r \rp}, \, \text{ where } G_X(U_{\r \rp}) = G_X[\r] U_{\r \rp} G_X^{\dagger}[\rp]
\end{equation}
Under a local gauge transformation $\widetilde{U_{\r \rp}} = W_{\r}U_{\r\rp}W_{\rp}^{\dagger}$, therefore
\begin{equation}
G_X \rightarrow \widetilde{G}_X = W_{\r} G_X W^{\dagger}_{X(\r)}
\end{equation}
We can use this gauge freedom to choose $G_{T_x} = \tau^{0}$. Now, consider the commutation of $T_x$ and $T_y$.  
\begin{eqnarray}
(G_{T_x} T_x) (G_{T_y} T_y) (G_{T_x} T_x)^{-1} (G_{T_y} T_y)^{-1}  = \eta_{T_x T_y} \tau^{0} \nonumber \\
\implies G_{T_y}(\r - \hat{x}) G_{T_y}^{-1}(\r) =  \eta_{T_x T_y} \tau^{0}
\end{eqnarray}
In an appropriate gauge, we can choose the solution as $G_{T_y}(x,y) = (\eta_{T_x T_y})^{x} \tau^{0}$. This choice of gauge, where both $G_{T_x}$ and $G_{T_y}$ are proportional $\tau^{0}$, is referred to as the uniform gauge \cite{WenSqLattice} as it preserves the translation invariance of SU(2) flux through any loop. 

Next, consider the commutations of time-reversal $\T$ with $T_x$ and $T_y$. We find that 
\beq
G_{\T}(\r - \hat{x})G_{\T}(\r)^{-1} = \eta_{\T T_x} \tau^{0} \, , \, G_{\T}(\r - \hat{y})G_{\T}(\r)^{-1} = \eta_{\T T_y} \tau^{0}
\eeq
Hence we can write the solution as $G_{\T}(x,y) =(\eta_{\T T_x})^x (\eta_{\T T_y})^y g_{\T}$, where $g_{\T} \in $SU(2). The added constraint $G_{\T}^2 = \eta_T \tau^{0}$ yields $ g_{\T}^2 =  \eta_T  \tau^{0}$. 

Let us consider the commutations of $P_x$ with $T_x,T_y$. 
\beq
(G_{P_x} P_x) (G_{T_x} T_x) (G_{P_x} P_x)^{-1} (G_{T_x} T_x) = \eta_{P_x T_x} \tau^{0} \implies G_{P_x}(\r) G_{P_x}(\r + \hat{x})^{-1} = \eta_{P_x T_x} \tau^{0} \nonumber \\
(G_{P_x} P_x) (G_{T_y} T_y) (G_{P_x} P_x)^{-1} (G_{T_y} T_y)^{-1} = \eta_{P_x T_y} \tau^{0} \implies G_{P_x}(\r) G_{P_x}(\r - \hat{y})^{-1} = \eta_{P_x T_y} \tau^{0}
\eeq
The solution is $G_{P_x}(x,y) = ( \eta_{P_x T_x})^{x} (\eta_{P_x T_y})^{y} g_{P_x}$, where $g_{P_x} \in$ SU(2) satisfies $g_{P_x}^{2} = \eta_{P_x} \tau^{0}$ since $G_{P_x}^{2} = \eta_{P_x} \tau^{0}$.

Similarly, for $P_y$ we find that $G_{P_y}(x,y) = ( \eta_{P_y T_x})^{x} (\eta_{P_y T_y})^{y} g_{P_y}$, where $g_{P_y} \in$ SU(2) satisfies $g_{P_y}^{2} = \eta_{P_y} \tau^{0}$ since $G_{P_y}^{2} = \eta_{P_y} \tau^{0}$.

Finally, we need to look at commutations of $P_x$ and $P_y$ with time-reversal $\T$, and between themselves. 
\beq
\label{fermConstraints}
(G_{P_x}P_x) (G_{\T} \T) (G_{P_x} P_x)^{-1} (G_{\T} \T)^{-1} = \eta_{\T P_x} \tau^{0} \implies g_{P_x} g_{\T} g_{P_x}^{-1} g_{\T}^{-1} = \eta_{\T P_x} \tau^{0}, \nonumber \\
(G_{P_y}P_y) (G_{\T} \T) (G_{P_y} P_y)^{-1} (G_{\T} \T)^{-1} = \eta_{\T P_y} \tau^{0} \implies g_{P_y} g_{\T} g_{P_y}^{-1} g_{\T}^{-1} = \eta_{\T P_y} \tau^{0}, \nonumber \\
(G_{P_x}P_x) (G_{P_y} P_y) (G_{P_x} P_x)^{-1} (G_{P_y} P_y)^{-1} = \eta_{P_x P_y} \tau^{0} \implies g_{P_x} g_{P_y} g_{P_x}^{-1} g_{P_y}^{-1} = \eta_{P_x P_y} \tau^{0}  .
\eeq
The full fermionic PSG on a rectangular lattice with time-reversal $\T$ is thus given by the following equations, together with the constraints set by Eq.~(\ref{fermConstraints}).
\begin{subequations}
\begin{align}
G_{T_{x}}(x,y) &= \tau^{0},
\\
G_{T_{y}}(x,y) &= (\eta_{T_xT_y})^{x} \tau^{0},
\\
G_{P_x}(x,y) &= ( \eta_{P_x T_x})^{x} (\eta_{P_x T_y})^{y} g_{P_x}, \; g_{P_x} \in SU(2), \; g_{P_x}^{2} = \eta_{P_x} \tau^{0},
\\
G_{P_y}(x,y) &= ( \eta_{P_y T_x})^{x} (\eta_{P_y T_y})^{y} g_{P_y}, \; g_{P_y} \in SU(2), \; g_{P_y}^{2} = \eta_{P_y} \tau^{0} ,
\\
G_{\T}(x,y) &= (\eta_{\T T_x})^x (\eta_{\T T_y})^y g_{\T}, \; g_{\T} \in SU(2), \; g_{\T}^{2} = \eta_{\T} \tau^{0}. \end{align}
\end{subequations}

\section{Trivial and non-trivial fusion rules}
\label{FusRules}
Consider a unitary symmetry operation $X^2 =1$ which is realized projectively on the anyons. To detect the symmetry fractionalization corresponding to $X$, we follow Ref.~\onlinecite{QiFu}. We act $X$ once on an excited state containing two anyons, whose positions are swapped by $X$. The symmetry action on an anyon is accompanied by additional gauge transformations, so we have
\begin{equation}
X \ket{a_{\r}} = U_{\r} \ket{a_{X(\r)}}, \; X \ket{a_{X(\r)}} = U_{X(\r)} \ket{a_{\r}}, \implies X^2\ket{a_{\r}} = U_{\r} U_{X(\r)} \ket{a_{\r}}
\end{equation}
Then, the phase factor we get on acting $X$ twice is given by $U_{\r} U_{X(\r)}$, which is nothing but $e^{i \phi_a}$, the phase corresponding to the anyon $a$.

First, consider acting $X$ on a physical wave-function $\ket{\Psi} = f_{\r}^{\dagger} f_{X(\r)}^{\dagger} \ket{G}$, with two fermionic spinons at $\r$ and $X(\r)$. Assuming that the ground state $\ket{G}$ is symmetric, we have
\begin{equation}
X \ket{\Psi} = (X f_{\r}^{\dagger} X^{-1})( X f_{X(\r)}^{\dagger} X^{-1}) \ket{G} =U_{\r} U_{X(\r)}  f_{X(\r)}^{\dagger} f_{\r}^{\dagger} \ket{G} = - U_{\r} U_{X(\r)} \ket{\Psi} = - e^{i \phi_f} \ket{\Psi} 
\end{equation}
This extra minus sign comes from reordering of the fermionic spinons under $X$, which is crucially dependent on the statistics of the fermion. 

Now, the same state can be thought of a pair of bound states of a bosonic spinon and a vison, i.e, 
\begin{equation}
\ket{\Psi} = b_{\r}^{\dagger} \phi_{\r}^{\dagger} b_{X(\r)}^{\dagger} \phi_{X(\r)}^{\dagger} \ket{G}.
\end{equation}
Applying $X$ on this state, there is no fermion reordering sign, and we get 
\begin{equation}
X \ket{\Psi} =  e^{i \phi_b}  e^{i \phi_v} \ket{\Psi}.
\end{equation}
Hence, comparing the two relations we find that in such cases, the fusion rule is non-trivial and carries an extra twist factor of $-1$, i.e,
\begin{equation}
 e^{i \phi_b}  e^{i \phi_v} = - e^{i \phi_f} .
\end{equation}

For the rectangular lattice, we want to figure out which symmetry fractionalization quantum numbers have non-trivial fusion rules. First, consider the reflections $P_x$ and $P_y$, and the inversion $I = P_x P_y$. All of these square to identity, implying the relations $P_x^2 =1,$ $P_y^2 = 1$ and $(P_x P_y)^2 = 1$ have non-trivial fusion rules. Now we use following the algebraic identity 
\beq
(P_x P_y)^2 = (P_x P_y P_x^{-1} P_y^{-1}) \cdot P_x^2 \cdot P_y^2 
\eeq 
Since the PSGs associated with $P_x^2$, $P_y^2$ and $(P_x P_y)^2$ have non-trivial fusion rules, the fusion rule for $P_x P_y P_x^{-1} P_y^{-1}$ must be non-trivial as well.

Next, note that the identity $P_x^{-1} T_x P_x T_x =1$ can also be written as $P_x^{-2}Y^{2} = 1$, where $Y = P_xT_x $. Now, $P_x^2$ and $Y^2$ both have non-trivial fusion rules, so the fusion rule for $P_x^{-1} T_x P_x T_x =1$ is trivial. Identical arguments show that $P_y^{-1} T_y P_y T_y =1$ has a trivial fusion rule.

\begin{figure}[!h]
\begin{center}
\includegraphics[scale=0.9]{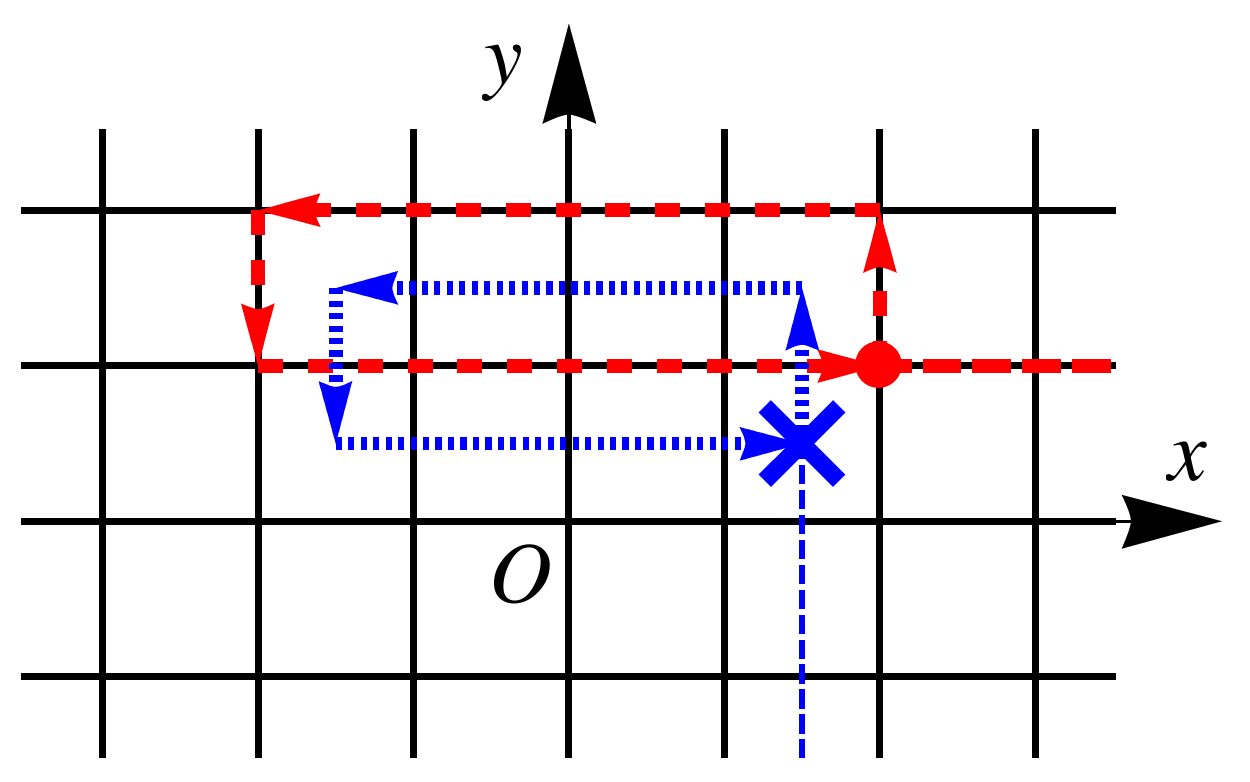}
\caption{Crossing of spinon (red blob) strings, represented by dashed red lines, and vison (blue cross) strings, represented by dotted blue lines, under $T_y P_x T_y^{-1} P_x^{-1}$}
\label{PxTyAnyonStrings}
\end{center}
\end{figure}

Now consider $P_x^{-1} T_y^{-1}P_x T_y$ and its counterpart $x \leftrightarrow y$. In this case, it is sufficient to act on single anyons, and we find that the spinon string has cut the vison string an even number of times under any of these operations, as illustrated in Fig. \ref{PxTyAnyonStrings}. Therefore, these commutation relations have a trivial fusion rule.  An analogous argument shows that $T_x^{-1} T_y^{-1} T_x T_y =1$ has a trivial fusion rule. 

Finally, let us consider time reversal symmetry. We know that both bosonic and fermionic spinons have half-spin with $\mathcal{T}^2 = -1$, whereas the vison is a spin-singlet with $\mathcal{T}^2 = 1$, so the fusion rule for $\mathcal{T}^2$ must be trivial. 

To derive the fusion rules of $R^{-1} \mathcal{T}^{-1} R \mathcal{T}$, where $R = P_x \text{ or } P_y$, we follow Ref. \onlinecite{Unification_Luetal}. We first consider the anti-unitary operator squared $(\mathcal{T} R )^2$. If we act $R^2$ on a pair of spinons and visons on the reflection axis, the spinon and vison strings cross. This implies that the phase picked up by a bosonic spinon relative to the bound state of a fermionic spinon and a vison, is $\pm i$ for the single reflection $R$. This is offset by the anti-unitary time reversal operator, which complex conjugates the wave function. Hence, the net relative phase is $(\pm i)^{*} \times (\pm i) = 1$, as illustrated in \cite{Unification_Luetal}. So, $(\mathcal{T} R )^2$ has a trivial fusion rule. Now, we use the algebraic identity
\begin{equation}
(\mathcal{T} R )^2 = (R^{-1} \mathcal{T}^{-1} R \mathcal{T} ) \cdot \mathcal{T}^2 \cdot R^2
\end{equation}
Since the PSGs associated with $\mathcal{T}^2$ and $(\mathcal{T}R)^2 $ have a trivial fusion rule, whereas that of $R^2$ obeys a non-trivial fusion rule, the PSGs of $R^{-1} \mathcal{T}^{-1} R \mathcal{T}$ must also have a non-trivial fusion rule. 

Finally, we consider the PSGs of $T_x^{-1} \mathcal{T}^{-1} T_x \mathcal{T}$. We again consider a similar setup as the previous case, with two spinons and two visons. Under $T_x$ followed by $T_x^{-1}$, there is no crossing of the spinon and vison strings - so there is no phase factor acquired by an indvidual bosonic spinon relative to the bound state of the fermionic spinon and the vison. Therefore, this commutation relation has a trivial fusion rule, and so does $T_y^{-1} \mathcal{T}^{-1} T_y \mathcal{T}$.

\section{Solution for the fermionic ansatz}
\label{Ferm1}
We need to find an ansatz $U_{\r \rp}$ such that $G_X X(U_{\r \rp}) = U_{\r \rp}$ for all symmetry operations $X$, where the gauge transformation $G_X$ corresponding to a symmetry operation $X$ has been derived from the fusion rules. Note that under time-reversal (slightly modified version as described in Ref. \onlinecite{WenSqLattice}), we have $\T(U_{\r \rp}) = - U_{\r \rp}$, so $g_{\T}$ must be non-trivial ($\neq \tau^{0}$) so that $G_{\T} \T(U_{\r \rp}) = U_{\r \rp}$, and therefore we require $\eta_T = -1$ for non-zero solutions. 
\begin{subequations}
\begin{align}
G_{T_{x}}(x,y) &= \tau^{0}
\\
G_{T_{y}}(x,y) &= (-1)^{x} \tau^{0}
\\
G_{P_x}(x,y) &=  g_{P_x},\; g_{P_x}^{2} =  \tau^{0} 
\\
G_{P_y}(x,y) &= (-1)^{x+y}  g_{P_y}, \; g_{P_y}^{2} = - \tau^{0} 
\\
G_{\T}(x,y) &=  g_{\T}, \; g_{\T}^{2} = - \tau^{0}
\end{align}
\end{subequations}
where the SU(2) matrices $g_{P_x}$, $g_{P_y}$ and $g_{\T}$ are satisfy the following (anti-)commutation relations.
\beq
[g_{P_x}, g_{\T}] = \{g_{P_y}, g_{\T} \} = [ g_{P_x}, g_{P_y}] = 0 
\eeq
In order to work with real hopping and pairing amplitudes in our ansatz, we follow Ref.~\onlinecite{Lu_Triangular} and choose $g_{\T}= i \tau^2$. Since $g_{P_x}$ commutes with both $g_{\T}$ and $g_{P_y}$, if $g_{P_y}$ is non-trivial, then $g_{P_x} = \tau^{0}$. We assume that this is the case, and choose $g_{P_y} = i \tau^{3}$ to get the solutions in Eq.~(\ref{FermGX}), also listed below:
\begin{subequations}
\begin{eqnarray}
G_{T_x}(x,y) &=& \tau^{0} ,\\
G_{T_y}(x,y) &=& (-1)^{x} \tau^{0} , \\
G_{P_x}(x,y) &=&  \tau^{0}, \\
G_{P_y}(x,y) &=& (-1)^{x + y} i \tau^{3}, \\
G_{\T}(x,y) &=& i \tau^{2} .
\end{eqnarray}
\label{FermGX1}
\end{subequations}
Note that $g_{P_y} = i \tau^{3}$ is a gauge choice, we could have as well chosen $g_{P_y} = i \tau^{1}$, or any properly normalized linear combination given by $g_{P_y} = i ( \text{cos} \theta \, \tau^{3} + \text{sin}\theta \, \tau^{1})$. However, all these choices lead to gauge-equivalent ansatz. Noting that $e^{i \theta \tau^2} \tau^{1} e^{-i \theta \tau^2} = \text{cos}(2\theta) \tau^{1} + \text{sin}(2\theta)\tau^{3}$, a mean-field matrix $U_{\r \rp}$ proportional to $\tau^{1}$ can be rotated to $\tau^{3}$ by a gauge transformation $W_{\r} = e^{i \theta \tau^2}$ with $\theta = \pi/2$. Therefore, we work with the first choice for convenience. 

First, we note from [\ref{Umatrix}] that $i U_{\r \rp} \in$ SU(2) upto a normalization constant in order to preserve spin-rotation symmetry, so we can expand in the basis of Pauli matrices as 
\beq
U_{\r \rp} = \sum_{\mu=0}^{3} \alpha^{\r \rp}_{\mu} \tau^{\mu}, \text{ where } i \alpha^{\r \rp}_{0}, \alpha^{\r \rp}_{1,2,3} \in \mathbb{R} 
\eeq 

$G_{\T} (U_{\r \rp}) = - U_{\r \rp} \implies \{U_{\r \rp}, \tau^{2}\} =0 \implies \alpha^{\r \rp}_{2} = 0$ for all bonds $\langle \r \rp \rangle$. Since the ansatz (not the spin-liquid) must break translational symmetry in the $y$ direction due to non-trivial $G_{T_y}$, we choose the following forms for the ansatz (upto third nearest neighbor):
\beq
U_{\r,\r + \hat{x}} = u_x (-1)^{y}, \, U_{\r,\r + \hat{y}} = u_y, \, U_{\r,\r + \hat{x} + \hat{y}} = (-1)^{y} u_{x+y}, \, U_{\r,\r - \hat{x} + \hat{y}} = (-1)^{y} u_{-x+y}, \, U_{\r,\r + 2\hat{x}} = u_{2x}, \, U_{\r,\r + 2\hat{y}} = u_{2y}. \nonumber \\
\eeq
Now we just apply the parity relations to each of the bonds in the ansatz. For the NN bonds
\beq
G_{P_x} P_x (U_{\r,\r + \hat{x}}) = U_{\r,\r + \hat{x}} \implies u^{\dagger}_{x} = u_{x}, \; G_{P_y}P_y (U_{\r,\r + \hat{x}}) = U_{\r,\r + \hat{x}} \implies \tau^{3} u_x \tau^{3} = - u_x, \nonumber \\
G_{P_x} P_x (U_{\r,\r + \hat{y}}) = U_{\r,\r + \hat{y}} \implies u_{y}= u_y, \; G_{P_y}P_y (U_{\r,\r + \hat{y}}) = U_{\r,\r + \hat{y}} \implies \tau^{3} u_y^{\dagger} \tau^{3} = - u_y .
\eeq
Together, these imply that $u_x = \Delta_{1x} \, \tau^{1}$ and $u_y = \Delta_{1y} \, \tau^{1}$ where both the pairing amplitudes are real. Similarly, we find that 
\beq
G_{P_x} P_x (U_{\r,\r + \hat{x} + \hat{y}}) = U_{\r,\r + \hat{x} + \hat{y}} \implies u_{-x+y} = - u_{x+y}, \; G_{P_y}P_y (U_{\r,\r + \hat{x} + \hat{y}}) = U_{\r,\r + \hat{x} + \hat{y}} \implies \tau^{3} u^{\dagger}_{-x+y} \tau^{3} = -u_{x+y} . \nonumber \\
\eeq Together, these imply that for the next-nearest neighbors
\beq
u_{x+y} = u_{-x+y} = \Delta_2 \tau^{1}.
\eeq
Analogous calculations show that the next to next nearest neighbors have a hopping term
\beq
u_{2x} = - t_{2x} \tau^{3}, \; u_{2y} = - t_{2y} \tau^{3}.
\eeq
One can also check that an on-site chemical potential term proportional to $\tau^{3}$ is allowed by the PSG. This ansatz describes a $\mathbb{Z}_2$ spin liquid, as it has both hopping and pairing terms for the fermionic spinons in any choice of gauge. 

Alternately, one can check that the SU(2) fluxes through different loops based at the same point are non-collinear, which also implies that the effective theory has a gauge group of $\mathbb{Z}_2$ \cite{WenSqLattice,LeeNagaosaWen}. Explicitly, consider the following two loops based at $\r$, $L_A: \, \r \rightarrow \r +  \hat{x} + \hat{y} \rightarrow \r + \hat{y} \rightarrow \r $ and $L_B: \, \r \rightarrow \vec{r} +  \hat{x} + \hat{y} \rightarrow \r -  \hat{x} + \hat{y} \rightarrow \r$. The product of $U_{\r \rp}$ on $L_A$ is proportional to $\tau^{1}$, whereas that on $L_B$ is proportional to $\tau^{3}$, which clearly point in different directions in SU(2) space.

\section{Alternative derivation of the specific fermionic PSG}
\label{alt-fPSG}
In this appendix, we present an alternative derivation of the fermionic PSG, which represents the same spin liquid state as the bosonic PSG in Eq.~\eqref{eq:bpsg} and Appendix~\ref{BosAnsPSG}. Instead of calculating the fractional quantum numbers of the fermionic spinon using the ones of the bosonic spinon and the vison, according to the fusion rules, here we derive this by directly mapping the bosonic mean-field wave function to a fermionic mean-field wave function, using the method introduced in the Supplemental Material of Ref.~\onlinecite{FanHongVBS2012}.

We start with the Schwinger-boson wave function in Eq.~\eqref{eq:bsrwf}, and we choose the weights to be $\xi_{\r\rp}=Q_{\r\rp}$ on the nearest-neighbor and next-nearest-neighbor bonds, and $\xi_{\r\rp}=0$ on other bonds, where the values of $Q_{\r\rp}$ are shown in Fig.~\ref{fig:ansatz}. With this choice, the wave function in Eq.~\eqref{eq:bsrwf} belongs to the phase described by the PSG in Appendix~\ref{BosAnsPSG}, because the wave function is invariant under the transformations in Eq.~\eqref{eq:bpsg}. We notice that although this wave function is constructed using the parameters of the mean-field Hamiltonian in Eq.~\eqref{eq:bsrmfh}, it is not the ground state of that Hamiltonian. However, it belongs to the same spin liquid phase as the ground state of that Hamiltonian.

Using the result in the Supplemental Material of Ref.~\onlinecite{FanHongVBS2012}, we can convert the Schwinger-boson wave function to the following Schwinger-fermion wave function,
\begin{equation}
  \label{eq:fsr-delta}
  |\Psi^f(s)\rangle=\sum_c s^{\delta_c}\prod_{(\r\rp)\in c}\zeta_{\r\rp}f_{\r\uparrow}^\dagger f_{\rp\downarrow}^\dagger|0\rangle,
\end{equation}
where $c$ runs over all possible nearest-neighbor and second-nearest-neighbor dimer coverings on the square lattice, 
$\zeta_{\r\rp}=\zeta_{\rp\r}$ are weights of the dimers, $\delta_c$ counts the number of dimer crossings in the covering, and each crossing contributes an extra weight factor $s$ to the wave function. With $s=-1$, the wave function $|\Psi^f(s=-1)\rangle$ exactly reproduces the Schwinger-boson wave function in Eq.~\eqref{eq:bsrwf}, if for every triangular plaquette $p$, the fermionic weights $\zeta_{\r\rp}$ satisfies
\begin{equation}
  \label{eq:bfloop}
  \prod_{(\r\rp)\in p}\zeta_{\r\rp}=-\prod_{(\r\rp)\in p}\xi_{\r\rp},
\end{equation}
where on the right hand side, the bonds are oriented in the counterclockwise direction. In other words, in each triangle, the flux in the fermionic model differs from the one in the bosonic model by $\pi$. One choice of weights satisfying Eq.~\eqref{eq:bfloop} is the following,
\begin{equation}
  \label{eq:fweights}
  \zeta_{\r,\r+\hat x}=(-1)^y Q_{(0,0)\rightarrow(1,0)},\quad
  \zeta_{\r,\r+\hat y}=Q_{(0, 0)\rightarrow(0, 1)},\quad
  \zeta_{\r,\r+\hat x+\hat y}=\zeta_{\r,\r-\hat x+\hat y}=(-1)^yQ_{(0,0)\rightarrow(1,1)}.
\end{equation}

The Schwinger-boson wave function can only be mapped to a wave function with a nontrivial weight of $s=-1$ for each pair of crossing bonds, which is different from the ordinary Schwinger-fermion wave function,
\begin{equation}
  \label{eq:3}
  |\Psi^f(s=+1)\rangle=\sum_c \prod_{(\r\rp)\in c}\epsilon_{\alpha\beta}\zeta_{\r\rp}f_{\r\alpha}^\dagger f_{\rp\beta}^\dagger|0\rangle
  =P_G\exp\left[\sum_{\r\rp}\zeta_{\r\rp}\epsilon_{\alpha\beta}f_{\r\alpha}^\dagger f_{\rp\beta}^\dagger\right]|0\rangle.
\end{equation}
 However, assuming that the two wave functions $|\Psi^f(s=\pm1)\rangle$ can be smoothly connected by varying $s$ from $-1$ to $+1$ (along the real axis), the two wave functions belong to the same phase, and the weights in Eq.~\eqref{eq:fweights} can be used to derive the fermionc PSG that constructs the same phase as the original bosonic PSG.

In particular, one can check that the wave function constructed using the weights in Eq.~\eqref{eq:fweights} is invariant under the lattice and time-reversal symmetries, if the fermionic spinon operator $f_{i\alpha}$ transforms according to the PSG in Eq.~\eqref{FermGX}.

We notice that this alternative derivation is not rigorous, as it depends on the assumption of the absence of any singularity in $|\Psi^f(s)\rangle$ when $s$ varies between $\pm1$. Nevertheless, this serves as a consistency check for the results presented in Sec.~\ref{sec:mapping}, without the explicit usage of the vison PSG and the fusion rules.

\section{PSG for the site bosons and constraints on $H_B$}
\label{siteBosPSG}
We derive the transformation of the boson-tuplet $B_{\r}$ under the projective transformations. We first focus on spatial symmetry operations $X_s$, which acts linearly (not projectively) on the $c$ fermion, and therefore all additional projective phase must come from the $f$ spinon. Recall that the $f$ fermion spinor transforms under a gauge transformation $G_X(\r)$ as 
\beq
\psi(\r) = \begin{pmatrix}
f_{\r \uparrow} \\
f^{\dagger}_{\r \downarrow}
\end{pmatrix} \rightarrow G_X(\r)\psi(\r).
\eeq
In our gauge choice, $G_{T_x} = G_{P_x} = \tau^{0}$, so these will just map $B_{\r}$ to itself. $G_{T_y}(\r) = e^{i\pi x} \equiv e^{- i \pi x}$ implies that $G_{T_y} B_{\r} = e^{i \pi x} B_{\r}$. Finally, we have 
\beq
G_{P_y}\psi_{\r} = e^{i \pi (x + y + 1/2)} \tau^{3} \psi_{\r} =e^{i \pi (x + y + 1/2)}   \begin{pmatrix} 1  & 0 \\ 0 & -1
\end{pmatrix}
 \begin{pmatrix}
f_{\r \uparrow} \\
f^{\dagger}_{\r \downarrow}
\end{pmatrix} = e^{i \pi (x + y + 1/2)}   \begin{pmatrix}
f_{\r \uparrow} \\
- f^{\dagger}_{\r \downarrow}
\end{pmatrix}.
\eeq
Therefore we see that under $G_{P_y}$, $f_{\r \sigma} \rightarrow e^{i \pi (x + y + 1/2)} f_{\r \sigma}$, and therefore $B_{\r} \rightarrow e^{i \pi (x + y + 1/2)} B_{\r}$. We conclude that the projective transformation under each spatial symmetry operation $X_s$ can be represented by just a phase $\phi_{X_s}$ on each boson, which we have listed in the main text in Eq.~(\ref{BosonPhases}).

Finally, we come to time-reversal, which acts non-trivially on both the $c$ and the $f$ fermions. Because on an additional gauge transformation $G_{\T} = i \tau^2$, we now have mixing between the two bosons. 
\beq
G_{\T} \T [\psi(\r)] = \begin{pmatrix} 0  & 1 \\ -1 & 0
\end{pmatrix}
 \begin{pmatrix}
f_{\r \uparrow} \\
- f^{\dagger}_{\r \downarrow}
\end{pmatrix} = \begin{pmatrix}
- f^{\dagger}_{\r \downarrow} \\
 - f_{\r \uparrow}
\end{pmatrix}.
\eeq
Therefore, we have $f_{\r \uparrow} \rightarrow - f^{\dagger}_{\r \downarrow}$, and $f_{\r \downarrow} \rightarrow - f^{\dagger}_{\r \uparrow}$, under time-reversal $\T$ combined with the gauge transformation $G_{\T}$. For the bosons, we find that
\beq
B_{1\r} &\rightarrow& \T(c^{\dagger}_{\r \uparrow}) G_{\T} \T (f_{\r \uparrow}) + \T(c^{\dagger}_{\r \downarrow}) G_{\T} \T (f_{\r \downarrow}) \nonumber \\
& = & c^{\dagger}_{\r \downarrow} (- f^{\dagger}_{\r \downarrow}) + (-c^{\dagger}_{\r \downarrow})(- f^{\dagger}_{\r \uparrow}) \nonumber \\
& = & \epsilon_{ \beta \alpha} f^{\dagger}_{\r \alpha} c^{\dagger}_{\r \beta} = B^{\dagger}_{2 \r} \,.
\eeq
and similarly, $B_{2\r} \rightarrow b^{\dagger}_{B \r}$. Imposing time-reversal symmetry on our hopping Hamiltonian in Eq.~(\ref{Hb}) therefore leads to the following constraints:
\beq
T^{11}_{\r \rp} = T^{22}_{\r \rp}, \; T^{12}_{\r \rp} = T^{21}_{\r \rp}.
\eeq
Notably, these constraints do not restrict these hoppings to take real values, and we can thus write down the hopping matrix as:
\beq
T_{\r \rp} = T_{\r \rp}^{d} \tau^0 + T_{\r \rp}^{od} \tau^{1},
\eeq
where $T^{d}$ and $T^{od}$ represent the diagonal and off-diagonal hopping matrix elements.

\bibliographystyle{apsrev4-1_custom}
\bibliography{flstar}

\end{document}